\documentclass[twocolumn,apj]{emulateapj}
\usepackage{relsize,cancel, graphicx, dcolumn,graphicx,color,booktabs,bbold, microtype,braket,mathrsfs,mathtools,simplewick,tikz,empheq,amssymb,amsmath,bm}
\usepackage{lineno}
\usepackage{soul}
\usepackage[colorlinks,plainpages=false,linkcolor=black,urlcolor=blue,citecolor=blue,pdfpagemode=UseNone,pdfstartview=FitH,allcolors=blue]
{hyperref}
\usepackage[T1]{fontenc}
\usepackage{cleveref}
\usepackage{amsmath}

\usepackage{tabularx}
\newcolumntype{L}{>{\raggedright\arraybackslash}X}
\newcolumntype{C}{>{\centering\arraybackslash}X}  

\begin{document}
\title{Exploring mutual information between IRIS spectral lines. II. Calculating the most probable response in all spectral windows.}
\author{Brandon Panos\altaffilmark{1,2}, Lucia Kleint\altaffilmark{1,2,3}}
\altaffiltext{1}{University of Applied Sciences and Arts Northwestern Switzerland, Bahnhofstrasse 6, 5210 Windisch, Switzerland}
\altaffiltext{2}{University of Geneva, CUI, 1205 Geneva, Switzerland}
\altaffiltext{3}{Leibniz-Institut f\"ur Sonnenphysik (KIS), Sch\"oneckstrasse 6, D-79104 Freiburg, Germany.}
\begin{abstract}
A three-dimensional picture of the solar atmosphere's thermodynamics can be obtained by jointly analyzing multiple spectral lines that span many formation heights. In paper I, we found strong correlations between spectral shapes from a variety of different ions during solar flares in comparison to the quiet Sun. We extend these techniques to address the following questions: which regions of the solar atmosphere are most connected during a solar flare, and what are the most likely responses across several spectral windows based on the observation of a single \ion{Mg}{2} spectrum? Our models are derived from several million IRIS spectra collected from 21 M- and X-class flares. We applied this framework to archetypal \ion{Mg}{2} flare spectra, and analyzed the results from a multi-line perspective. We find that (1) the line correlations from the photosphere to the transition region are highest in flare ribbons. (2) Blue shifted reversals appear simultaneously in \ion{Mg}{2}, \ion{C}{2} and \ion{Si}{4} during the impulsive phase, with \ion{Si}{4} displaying possible optical depth effects. \ion{Fe}{2} shows signs of strong emission, indicating deep early heating. (3) The \ion{Mg}{2} line appears to typically evolve a blue-shifted reversal that later returns to line center and becomes single peaked within 1-3 minutes. The widths of these single peaked profiles slowly erode with time. During the later flare stages, strong red wing enhancements indicating coronal rain are evident in \ion{Mg}{2}, \ion{C}{2}, and \ion{Si}{4}. Our framework is easily adaptable to any multi-line data set, and enables comprehensive statistical analyses of the atmospheric behavior in different spectral windows.
\end{abstract}

\section{Introduction}
Flares affect many heights throughout the solar atmosphere. According to the standard flare model \citep[e.g.][and references therein]{standard_model2, Arnold}, they are initiated by the reconnection of field lines in the corona, where magnetic energy is converted into kinetic energy through the acceleration of particles, both towards the lower solar atmosphere and into space. As the accelerated particles precipitate through denser atmospheric layers towards the solar surface, they lose their energy via several mechanisms, with a significant fraction of this energy being converted into radiation, heat, and bulk particle motions. Many spectral lines respond to the evolving atmosphere by changing their shapes, intensities, and Doppler shifts.

Although the intensity of certain lines may carry vital information about the thermal structure of the atmosphere,  \citep[e.g.][]{LeenaMgII_qs2,Temp_mg_2018}, it is often the case that the majority of the atmospheric variation is encoded within the line shapes themselves. For instance, the width of optically thin lines such as \ion{O}{1} 1355.6 $\text{\AA}$ are excellent diagnostics of non-thermal chromospheric velocities \citep{Hsiao_OI}. The ratios between line pairs where at least one of the lines is either populated by, or decays from a metastable state, can be used to derive electron density estimates \citep[e.g][]{Munro,Flower,general_density_methods}, with additional uncertainties being mitigated when the lines come from the same ion and have the same or little temperature dependence, such as the IRIS observed \ion{O}{4} intersystem multiplets around 1400 $\text{\AA}$. In some cases, even after normalizing out the intensity information of the entire solar spectrum, it is still possible to extract temperature information from the width of particular lines or carefully selected line ratios \citep{Heroux_temp_from_ratios}.

Analyzing the shape of a single spectral line in isolation can however be misleading, since a degenerate number of thermodynamic conditions (changing densities, temperatures, or velocities) can lead to the same spectral shape \citep{Rubio_da_Costa_2017}. A promising approach to resolve these ambiguities is to analyze multiple spectral lines with different formation heights simultaneously, providing the complementary information necessary to restrict the thermodynamic solution space, and reconstruct the behavior of the solar atmosphere during flares, via multi-line inversions \citep[e.g.][]{Inversions}.

In addition to studying multiple line shapes simultaneously, there are strong motivations for extending this practice to the study of multiple flares. At present, the large majority of flaring literature is predicated on single flare studies. Although particular flares are of interest, the development of a general theoretical framework relies on the identification of commonalities, both thermodynamic and topological in nature, across all flares \citep[see for example][]{Panos_2018,Chrom_Evap_CII_FeXXI}.

In Paper I \citep{Paper1} of this two paper series, we combined the above two guiding philosophies in a multi-line, multi-flare analysis that investigated the correlations between spectral shapes from different ions, both under quiet Sun and flaring conditions. The linear and non-linear correlations were distilled into an information theoretic quantity introduced by \cite{Shannon_MI}, called the mutual information (MI), and calculated using two independent machine learning techniques. Both techniques returned complementary results, showing significant enhancements in correlations for flaring atmospheres in comparison to their quiet Sun counterparts. These results were interpreted as a sign of strong three-dimensional couplings within flaring atmospheres. While the result indicated the degree to which certain spectral lines were correlated, the MI did not provide us with an understanding of how these correlations were distributed amongst the individual spectra.

In this second paper, we will introduce techniques that promote the analysis of correlations on a spectrum-to-spectrum basis, allowing us to answer questions such as: given an observed spectrum from one line, which spectral shapes are most likely to be observed in another spectral window formed over a different range of heights? This technique will be used to analyze a few key \ion{Mg}{2} flare spectra, with the aim of resolving some open questions about their formation properties.\\

\section{Data}
\label{Data_preprocessing}
We used the same flare data set from Paper I, which consists of 21 large flare observations by the IRIS satellite from the years 2014-2015. The list of flare observations is repeated from Paper~I in Table~\ref{obs} for reference, and contains a variety of observational modes, cadences, and slit orientations. The restriction to large M- and X-class flares, with the IRIS slit positioned directly over the flare ribbon, is advantageous for our analysis of correlated flaring profiles. Each IRIS observation also covers several spectral windows, which are explained in the next section. We manually selected the time range of each observation to only include the flaring period. Since the spectrograph's field-of-view ($130\times 175~\text{arcsec}^2$) is comparatively large with respect to an average active region, and because the activity of the Sun dynamically changes over the course of an observation, many non-flaring spectra are captured by the spectrograph. In-order to restrict our study to spectra commonly observed outside of quiet Sun conditions, we trained a \textit{variational autoencoder} on \ion{Mg}{2} quiet Sun profiles and used the reconstruction error to dynamically filter quiet Sun components (see section 2.1 of Paper~I). This resulted in a primary data set consisting of $4788392$ spectra $\times~5$ spectral regions, about 50 \% of the original data set. This filtering was done independently for each pixel along the slit and each time step, meaning that a given pixel was considered only for the time range during which the spectral shape did not look like a typical quiet Sun spectrum.

In order to make the data machine learning compatible, a number of preprocessing steps were taken in addition to the already pre-processed IRIS level2 data: 1) Spectra with missing data and large negative values were removed. 2) Likewise, overexposed spectra with more than 5 consecutive wavelength points with the same intensity were removed. 3) The variety of spectral resolutions across observations were homogenized by interpolating the spectra onto a common wavelength grid, such that the number of wavelength points are allowed to differ between lines but not within the same line. For instance, spectra for the \ion{Mg}{2} line were spline-interpolated onto a 240 wavelength grid, while all spectra across observations for \ion{Fe}{2} were interpolated to a fixed grid size of 100. In the machine learning context, this step is vital and ensures that the features are of the same dimensions and therefore carry the same precedence. 4) We finally normalized each spectrum by its maximum value, which places the emphasis on spectral shape rather than the otherwise dominant intensity. 

The pipeline's output returns several clean flaring data sets, one for each spectral region. The analysis of this study was then performed on pairs of data sets ($\mathcal{L}i| \mathcal{L}j$), where $\mathcal{L}i$ and $\mathcal{L}j$ correspond to the data sets of spectral lines $i$ and $j$, respectively. We note that \ion{Si}{4} is often overexposed during flares, and therefore blindly dropping spectra across all ions would result in a sparse data set.  In order to mitigate this issue, we treated each pair of spectral lines separately, so that if the $k$'th \ion{Si}{4} spectrum was overexposed, the corresponding index in ($\mathcal{L}i$| \ion{Si}{4}), where $\mathcal{L}i$ refers to any other spectral line, would be dropped, however, the spectra corresponding to this index in (\ion{Mg}{2}|\ion{C}{2}) for instance, would be retained.

\begin{table}[t]
\caption{Flare observations}\centering 
\begin{tabularx}{.45\textwidth} { >{\hsize=.1cm\raggedright\arraybackslash}X >{\centering\arraybackslash}X >{\centering\arraybackslash}X>{\centering\arraybackslash}X>{\raggedleft\arraybackslash}X }
\toprule\toprule
\#& Class&Date&Time Obs Start&OBSID\\ 
\midrule
1&M1.0& 2014-06-12&11:09&3863605329\\
2&M1.0&2014-11-07&09:37&3860602088 \\
3&M1.1&2014-06-12&18:44&3863605329\\
4&M1.1&2014-09-06&11:23&3820259253\\
5&M1.1&2015-08-21&16:01&3660104044\\
6&M1.3&2014-10-26&18:52&3864111353\\
7&M1.4&2015-03-12&05:45&3860107053 \\
8&M1.8&2015-03-11&04:46&3860259280\\
9&M2.3&2014-11-09&15:17&3860258971\\
10&M2.9&2015-08-27&05:37&3860605380\\
11&M3.4&2014-10-27&20:56&3864111353\\
12&M3.9&2014-06-11&18:19&3863605329\\
13&M6.5&2015-06-22&17:00&3660100039\\
14&M7.3&2014-04-18&12:33&3820259153\\
15&M8.7&2014-10-21&18:10&3860261353\\
16&X1.0&2014-03-29&14:09&3860258481\\
17&X1.6&2014-09-10&11:28&3860259453\\
18&X1.6&2014-10-22&08:18&3860261381\\
19&X2.0&2014-10-27&14:04&3860354980\\
20&X2.1&2015-03-11&15:19&3860107071\\
21&X3.1&2014-10-24&20:52&3860111353\\
\bottomrule
\label{obs}
\end{tabularx}
\end{table}

\begin{figure*}[tbh] 
\centering
\includegraphics[width=.95\textwidth]{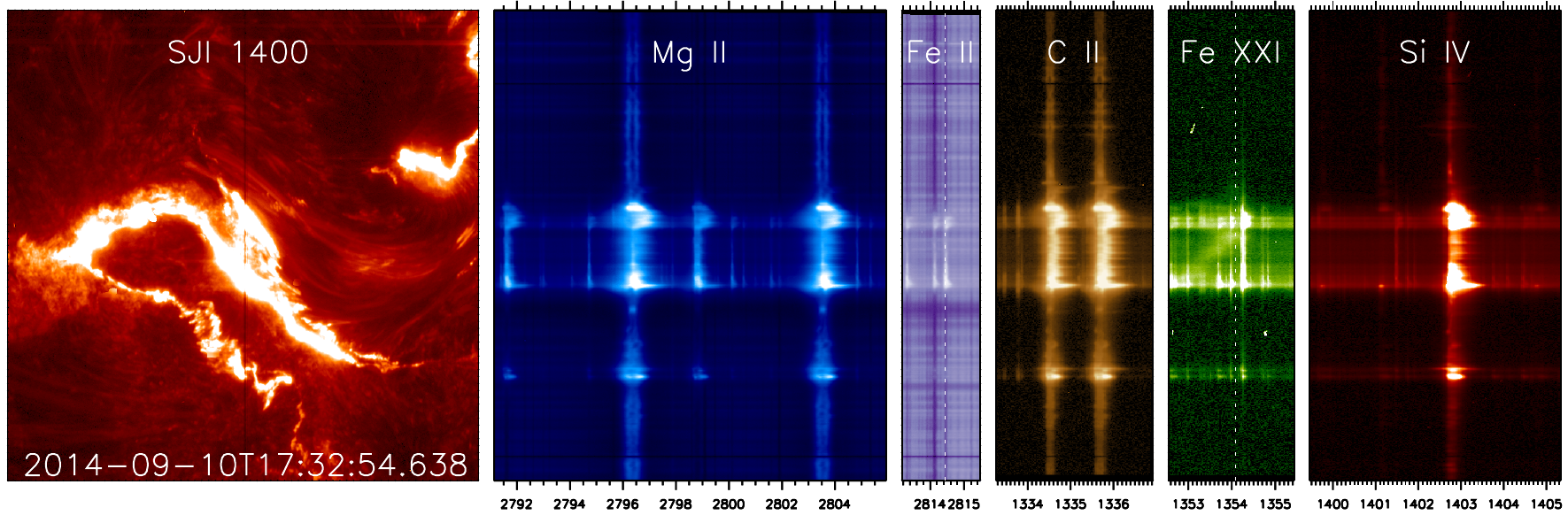}
\caption{Overview of the spectral windows used for the analysis in this paper. All images and spectra were scaled to the power of 0.3 for better contrast. Different spectral shapes are visible along the slit (vertical axis), in particular, upflows of different velocities in \ion{Fe}{21} and strong emission at the location of the flare ribbons, which also partially show asymmetries towards the red wavelengths (downflows). A movie showing the evolution of these flare spectra is available online.}
\label{overview}
\end{figure*}

\subsection {Spectral lines}
In this section, we present an overview of some of the most important IRIS spectral lines, whose couplings we will investigate during the course of this paper. \Cref{overview} shows an example of the selected spectral windows and the many different spectral shapes that can occur during the course of a solar flare. The different spectral windows are showcased to the right of the solar image (SJI 1400), and are collected by IRIS' spectrograph whose slit can be seen as a black vertical line in the left most image. The aim of this research is to predict the most statistically likely responses within the \ion{Fe}{2}, \ion{C}{2}, \ion{Fe}{21} and \ion{Si}{4} spectral windows, given only the \ion{Mg}{2} spectra in blue. These methods are not confined to this particular arrangement, and can be applied with any spectral window as a basis.\\

\textbf{\ion{Mg}{2}:}
The \ion{Mg}{2} h\&k resonant doublets are amongst the strongest lines in the solar UV spectrum, with wings that follow the temperature structure of the photosphere and peaks that form in the mid-chromosphere. On account of the high relative abundance of magnesium \citep{Composition}, the core formation height extends into the upper chromosphere, where radiative de-excitation exceeds collisional de-excitation by a factor of four \citep{MgII_qs1}, resulting in a loss of temperature sensitivity in the source function and ubiquitous quiet Sun central reversals. 

Since the lines are strongly scattering and form in a relatively low density environment, the velocity field remains unconfused in the frame of the atom, and partial frequency distribution (PRD) becomes important \citep{PRD}. Including PRD effects into radiative transfer codes ordinarily results in lower inner wing intensities and higher peak intensities with cores that are relatively well approximated by complete redistribution (CRD) \citep{MgII_qs1}. The intensity ratios of the \ion{Mg}{2} h\&k lines can be used as an opacity diagnostic, with a k/h-ratio of 2:1 and 1:1 indicating optically thin and thick line formation, respectively. The \ion{Mg}{2} h\&k  lines provide excellent upper and mid chromospheric diagnostics of temperatures and velocities \citep{LeenaMgII_qs2}. During solar flares, spectral profiles are greatly enhanced and broadened, with significant Doppler shifts resulting in a multitude of spectral shapes \citep{Kerr_1, rubiodacostaetal2016, Panos_2018, Mg_ran1, multi_line_flare}. The increased densities allow the line cores to re-couple to the Planck function and serves as one possible mechanism for the formation of single peaked spectra. \cite{Kerr_mgII_1} showed that PRD effects are still important even in the presence of increased collisional rates from enhanced densities. Furthermore, the assumption of statistical equilibrium appears to hold for strong flares \citep{Kerr_mgII_2}. The \ion{Mg}{2} k-line also has a companion of triplets at 2791.60, 2798.75 and 2798.82 \AA, with the two mixed subordinate lines in the red wing going into emission during flares. \cite{Zhu_2019} found that during solar flares, the formation height of the subordinate lines move from the lower to the upper chromosphere, and should therefore not be used as upper photospheric diagnostics.\\

\textbf{\ion{C}{2}:}
So far the \ion{C}{2} lines have only been modeled for the quiet Sun \citep{Adam2}, and not the flaring atmosphere. The \ion{C}{2} multiplet consists of  two strong resonant lines at 1334.53 $\text{\AA}$ and 1335.71 $\text{\AA}$, with a weak blend at 1335.66 $\text{\AA}$. Together they represent the strongest lines within the IRIS FUV1 passband, and one of the few optically thick lines that form around the base of the transition region \citep{CII_qs1}. The \ion{C}{2} lines share many characteristics with \ion{Mg}{2} h\&k, offering equivalent, if not slightly degraded diagnostics, which are nevertheless complementary. For the case of the quiet Sun, ionization equilibrium is controlled for low temperatures via photoionization and radiative recombination (\ion{C}{1}/\ion{C}{2}) and  by collisional ionization and dielectronic recombination for higher temperatures (\ion{C}{2}/\ion{C}{3}) \textcolor{blue}. Horizontal scattering effects are also important for the line core intensity. Depending on how much matter is at temperatures between 14 -- 50kK, the \ion{C}{2} line cores can form just above or below the formation height of \ion{Mg}{2}, and are well modeled by CRD \citep{CII_qs1}. Similar to \ion{Mg}{2}, 3D scattering effects play an important role within the line cores, which makes the formation properties strictly non-local \citep{CII_qs1}. The line ratios can be used as a complementary, but much weaker diagnostic of opacity than magnesium, with ratios other than 1.8 indicating optically thick emission, while ratios precisely at 1.8 remain ambiguous. The main distinction from \ion{Mg}{2} is that \ion{C}{2} has a narrower formation range and contains more optically thin emission.\\

\textbf{\ion{Si}{4}}: {The \ion{Si}{4} resonance lines at 1393.75 and 1402.77 $\text{\AA}$ are some of the strongest lines in IRIS' FUV channel, and provide insight into the structure and dynamics of the transition region. They form outside of statistical equilibrium \citep{Olluri_2015, Mart_2016} and are often assumed to be optically thin, although this assumption may not always be valid for flaring atmospheres. Significant deviations from the optically thin line ration of 2 (the ratio of the resonant lines spontaneous radiative decay rates) have been noted both in observations \citep[e.g.,][]{Bartoe_1975,Brannon_2015} and simulations \citep{Thick_SiIV}. The ratio between the \ion{Si}{4} and spin-forbidden \ion{O}{4} lines have been suggested as a potential diagnostic tool for probing high density regimes, ($N_e\sim 10^{13}\text{cm}^{-3}$), synonymous with flaring atmospheres, however, several issues regarding differences in peak formation temperatures, and ambiguities in the fractional ion abundances bring into question the validity of these results \citep{Judge_2015}. Additionally, some \ion{Si}{4} lines are blended with the \ion{O}{4} lines, which become particularly problematic for very large \ion{Si}{4} emissions during flares.} \\

\textbf{\ion{Fe}{2}:}
The literature for the weak lower chromospheric \ion{Fe}{2} 2814.445  $\text{\AA}$ line is very sparse, but has recently received some attention as a potential flaring diagnostic \citep{Adam2}. Furthermore, the line never saturates in IRIS observations \citep{Condensation}, offering continual diagnostic coverage even over the brightest flare kernels. \citet{Adam1} showed that a 5F11 high energy beam model produces two flare regions, a stationary and condensation layer moving at approximately 20-55 km s$^{-1}$. This model helps explain the observed red wing asymmetry over bright flare kernels and flare footpoints. These enhanced red wing components are endemic to chromospheric lines in general during the impulsive flare phase, and are a direct signature of strong chromospheric condensation first modeled by \citet{Fisher}. Since the \ion{Fe}{2} line is optically thinner than most chromospheric flare lines (but thicker than the NUV continuum emission), the hope is that \ion{Fe}{2} can be used to restrain flare induced velocity fields at large column mass, however, at present, there are several notable discrepancies between models and observations, some of which pertain to intensity ratios, and others with regards to the temporal evolution of the red wing \ion{Fe}{2} components. High cadence IRIS observations from \citet{Condensation} have highlighted several of these temporal disagreements. The observations always appear more gradual than their simulated counterparts, which show immediately comparable core and red wing intensity enhancement at beam onset (within a few seconds for the red wing), accompanied by a rapidly diminishing red wing enhancement towards its stationary position within about 10 seconds. The observations on the other hand show that the onset of the red wing enhancement is gradual, taking about $30$ s to be comparable in intensity to the rest wavelength component, and that a similar length of time (rather than the simulated $10$s) is required for the satellite to relax back to its stationary position.\\

\textbf{\ion{Fe}{21}:} Other than the often very weak \ion{Fe}{12} line, the 10 MK coronal \ion{Fe}{21} 1354.08 $\text{\AA}$ forbidden line is the only coronal line in IRIS' diagnostics. It is blended primarily with the strong \ion{C}{1} 1354.288 $\text{\AA}$ line, which is present even outside of flaring regions, as well as a few less significant narrow \ion{Fe}{2} line blends. Despite the low emissivity of \ion{Fe}{21}, it is clearly observed during solar flares in locations that exceed temperatures of 10 MK, these include the energy release site, flare ribbons and post flare loops. The \ion{Fe}{21} line usually first appears in flare ribbons, strongly blue-shifted with upflow velocities in excess of 200 km $\text{s}^{-1}$, in agreement with models of chromospheric evaporation, and decays to quasi-stationary within about 10 minutes \citep{innesetal2003,LoopFootBlueGraham, youngetal2015}. Downflows appear subsequently, at velocities below a few tens km $\text{s}^{-1}$ \citep{battagliaetal2015,Tian_2014}. Although \ion{Fe}{21} has provided crucial model constraints for chromospheric evaporation, the physical mechanism responsible for the nonthermal >100 km $\text{s}^{-1}$ widths at flare footpoints remains elusive. A recent study by \cite{Polito_2019}  used hydrodynamic models to dismiss the most popular explanation of sub-resolution plasma flows at different velocities, showing that an additional physical mechanism, such as plasma turbulence or very large ion temperatures is necessary for a complete explanation. A speculative dismissal of the superposition scenario was already put forward several decades earlier by \cite{Antonucci_1986}, who noted that it would be unlikely for up- and down-flowing material to exactly balance each other to produce the archetypal symmetric \ion{Fe}{21} profile commonly observed. This line is still actively challenging our understanding of energy transport within solar flares.

\section{Point-wise correlation measures}
\label{Point_Wise}
In this section, we refresh some of the concepts encountered in the first paper, and introduce a new quantity called the \textit{Point-Mutual-Information} or PMI, which allows us to investigate the correlation between lines on a spectrum-to-spectrum basis. We then explain how the PMI can be extracted from the numerical method discussed in the first paper, and use the results to confirm the hypothesis that spectra located directly over the flare ribbons contribute most to the inter-line correlations.

\subsection{Mutual and Pointwise information}
In Paper I of this series, we discussed the use of Mutual Information (MI) as a measure for the correlation between the shapes of different spectral lines. This quantity was given by
\begin{equation}
\text{MI}(\mathcal{L}1 ; \mathcal{L}2)=\sum_{x\in\mathcal{L}1}\sum_{y\in\mathcal{L}2} p(x, y) \log \frac{p(x, y)}{p(x) p(y)},
\label{mutual_information}
\end{equation}
and can be calculated using either a categorical method involving the k-means algorithm of \cite{macqueen1967}, or  a numeric method using a neural network called a Mutual Information Neural Estimator (MINE) \citep{MINE_2018}. The variables $x$ and $y$ represent spectra from lines $\mathcal{L}1$ and $\mathcal{L}2$, with the term $p(x,y)$ called the joint probability distribution, describing the chance of observing two specific types of spectra simultaneously, while the marginal probabilities $p(x)$ and $p(y)$ tell us how frequent each spectral type is. In-order to measure the joint and product of the marginal probabilities, we either sample spectra for both lines $\mathcal{L}1$ and $\mathcal{L}2$ from the same pixel, or in the latter case, from different pixels. In practice, $x$ and $y$ are not the actual spectra, but a set of reduced and highly relevant features, automatically derived by a neural network. These features could be a combination of quantities such as line widths, Doppler shifts and asymmetries. To calculate the correlation between any two spectral lines, say \ion{Mg}{2} and \ion{C}{2}, one has to perform the following procedure: 1)  reformulate Eq.(\ref{mutual_information}) to make it amenable to variational maximization techniques, 2) obtain a large reservoir of both \ion{Mg}{2} and \ion{C}{2} spectra (in our case several million), 3) derive the salient features of each line to serve as inputs $x$ and $y$, 4) calculate the joint and marginal probabilities by sampling spectra from the same and different pixels respectively. As it turns out, back propagation function approximators such as the MINE-network, allow us to automatically estimate the aforementioned probabilities and features while simultaneously solving an optimization problem (for an extensive description of this method please refer to Paper I). This results in a single value whose magnitude describes the degree to which both lines are correlated. If this value is high, then there is a good possibility that we could predict the shape of a \ion{C}{2} spectrum, just by knowing the shape of the \ion{Mg}{2} spectrum within the same pixel. On the other hand, because Eq.(\ref{mutual_information}) is an aggregate of many pairs of spectra, a low MI does not necessarily rule out the possibility of the existence of a small subset of strongly correlated spectra.

\begin{figure}[t] 
\centering
\includegraphics[width=.4\textwidth]{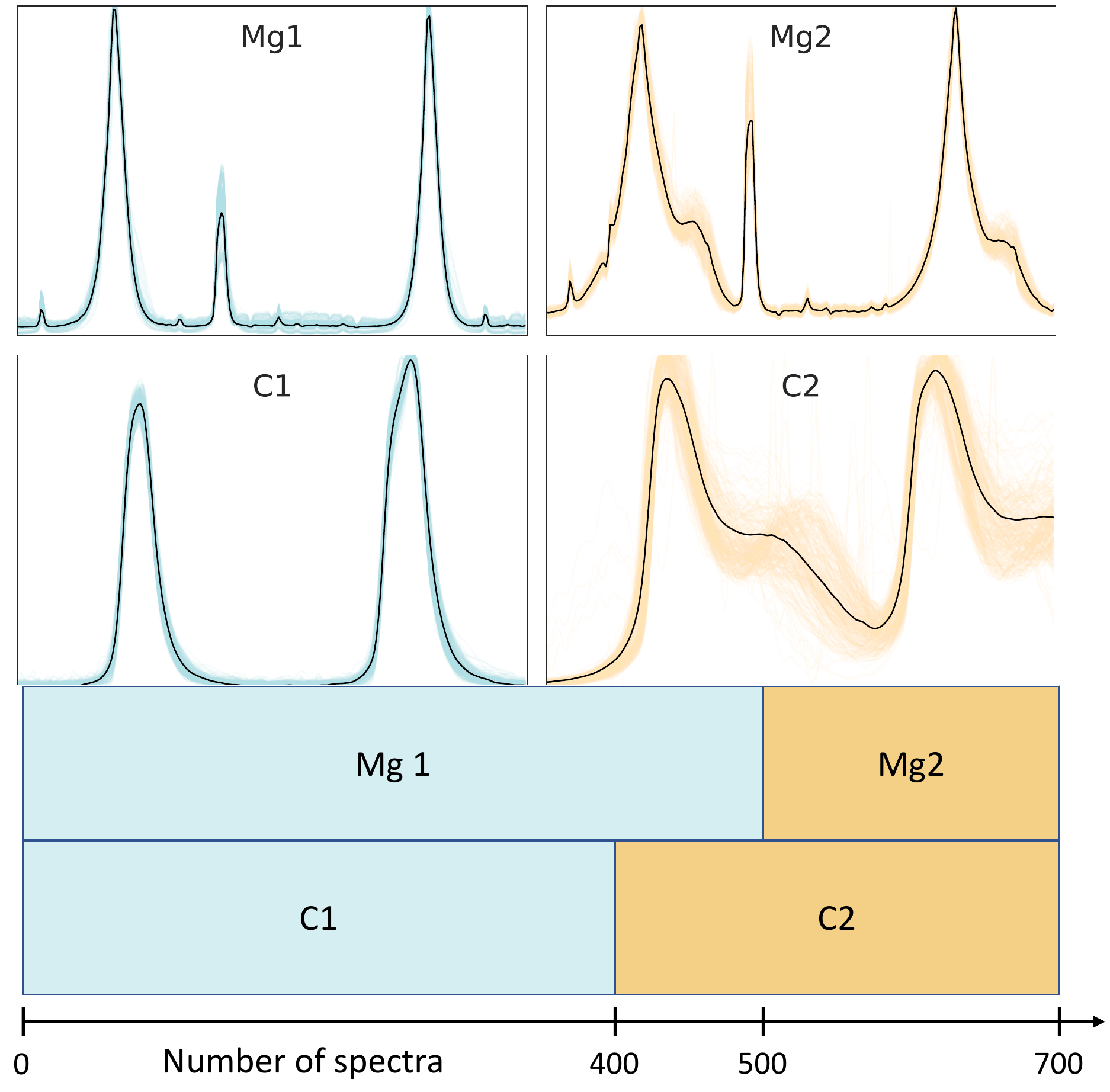}
\caption{Collection of synthetic data to test the convergence properties of PMI from our trained model $T_\theta$. Each of the 4 panels represents a different group found by k-means. The group centroids are displayed in black, while each groups members are overplotted in color. The number of spectra from each group as well as their pixel alignments can be seen in the lower colored bars. For example: There are 500 Mg1 type spectra connected with 400 C1 spectra, as well as 200 Mg2 type spectra connected with 300 C2 spectra. Controlling the ratio of groups and the connections between them allows us to calculate the true PMI between each spectrum, which in turn provides us with a benchmark for testing the convergence properties of our model.}
\label{pmi_synth}
\end{figure}

To address this uncertainty, instead of deriving the aggregate MI, the summand 
\begin{equation}
\text{PMI}(x;y) \triangleq \log\frac{p(x,y)}{p(x)p(y)},
\label{PMI1}
\end{equation}
of Eq.(\ref{mutual_information}), called the Pointwise-Mutual-Information (PMI), allows us to understand how the correlations are distributed among the individual spectra, and which spectral shapes are responsible for the largest contribution to the MI. The PMI is the ratio of the probability that two spectra from different lines occur simultaneously $p(x,y)$, with what we would expect from chance $p(x)p(y)$. As an example: Imagine that we have a photospheric line \ion{Fe}{2}, and a transition region line \ion{Si}{4}. If 90\% of the spectra from the photospheric line are indistinguishable absorption profiles and only 10\% of the profiles are in emission, then the fact that a specific type of profile shape from the \ion{Si}{4} line always appears with the photospheric absorption profiles is not statistically significant, because random chance would dictate this to be the case. On the other hand, if the same transition region spectra always occur with photospheric spectra in emission (but never with those in absorption), regardless of their 10\% sparsity, then the confidence that the spectral shapes are correlated is statistically more significant. This common sense deduction is precisely what PMI captures. The range of allowed values is given by the interval
\begin{equation}
-\infty < \text{PMI}(x ; y) \leq \min [-\log p(x),-\log p(y)],
\end{equation}
with larger positive numbers corresponding to stronger correlations. We speculate that the resultant hypothesis function of a carefully trained and converged MINE-network should be able not only to optimize over large aggregates of spectra, but also on a spectrum-to-spectrum basis. 

\begin{figure}[!tb] 
\centering
\includegraphics[width=.49\textwidth]{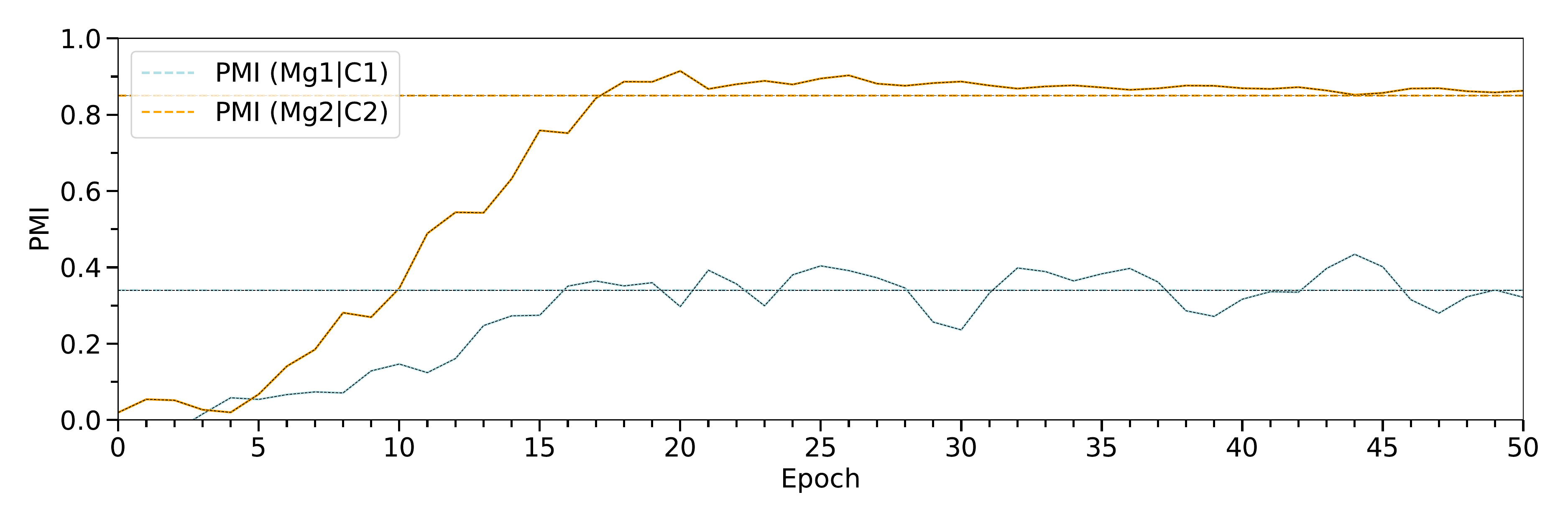}
\caption{Convergence of PMI between spectra from group Mg1 and C1 (blue curve) and Mg2 and C2 (orange curve). The curves were constructed by calculating the PMI between the centroids (not the raw spectra)  in Figure \ref{pmi_synth}, at each instance of the models training. For example: At Epoch 10, the PMI is underestimated in both cases because the model has only been able to update its internal parameters 10 times, leading to a suboptimal model. The horizontal orange and blue lines represent the true PMI calculated from the ratios given in Figure \ref{pmi_synth}. We see that the estimated PMIs quickly converge to their true values. Importantly, this is true when using centroids instead of raw data.}
\label{convergence_of_pmi}
\end{figure}

\begin{figure*}[!th] 
\centering
\includegraphics[width=.33\textwidth]{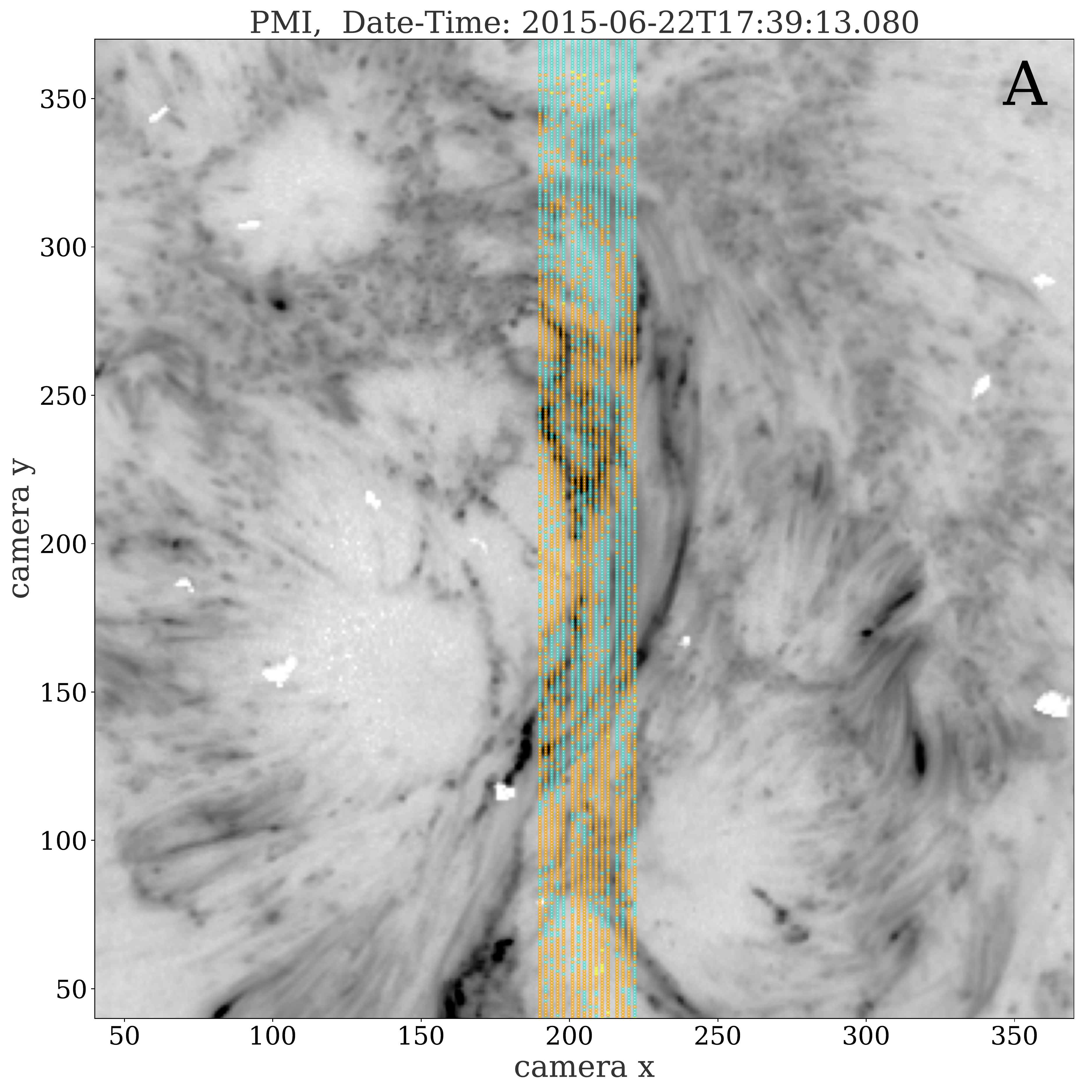}
\includegraphics[width=.33\textwidth]{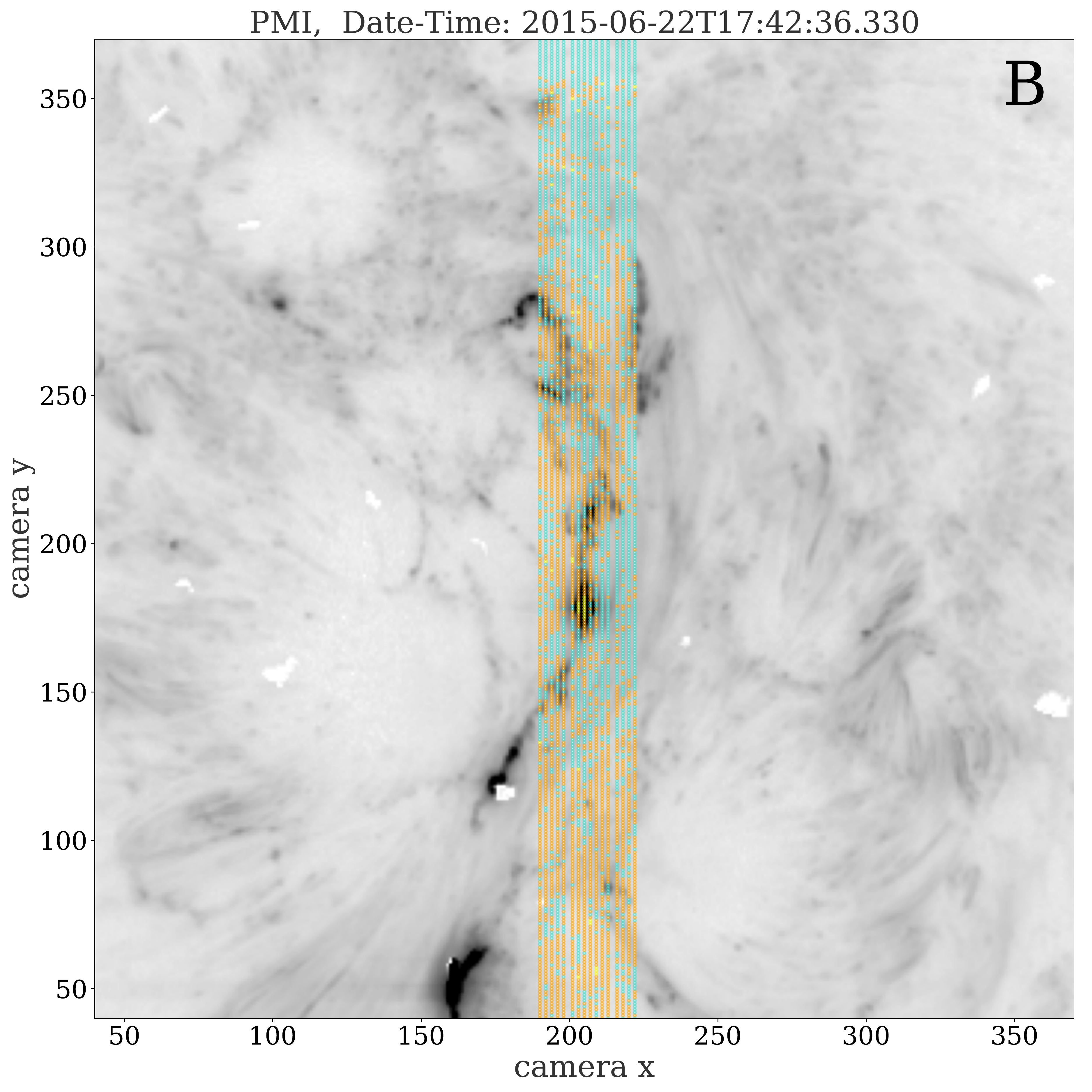}
\includegraphics[width=.33\textwidth]{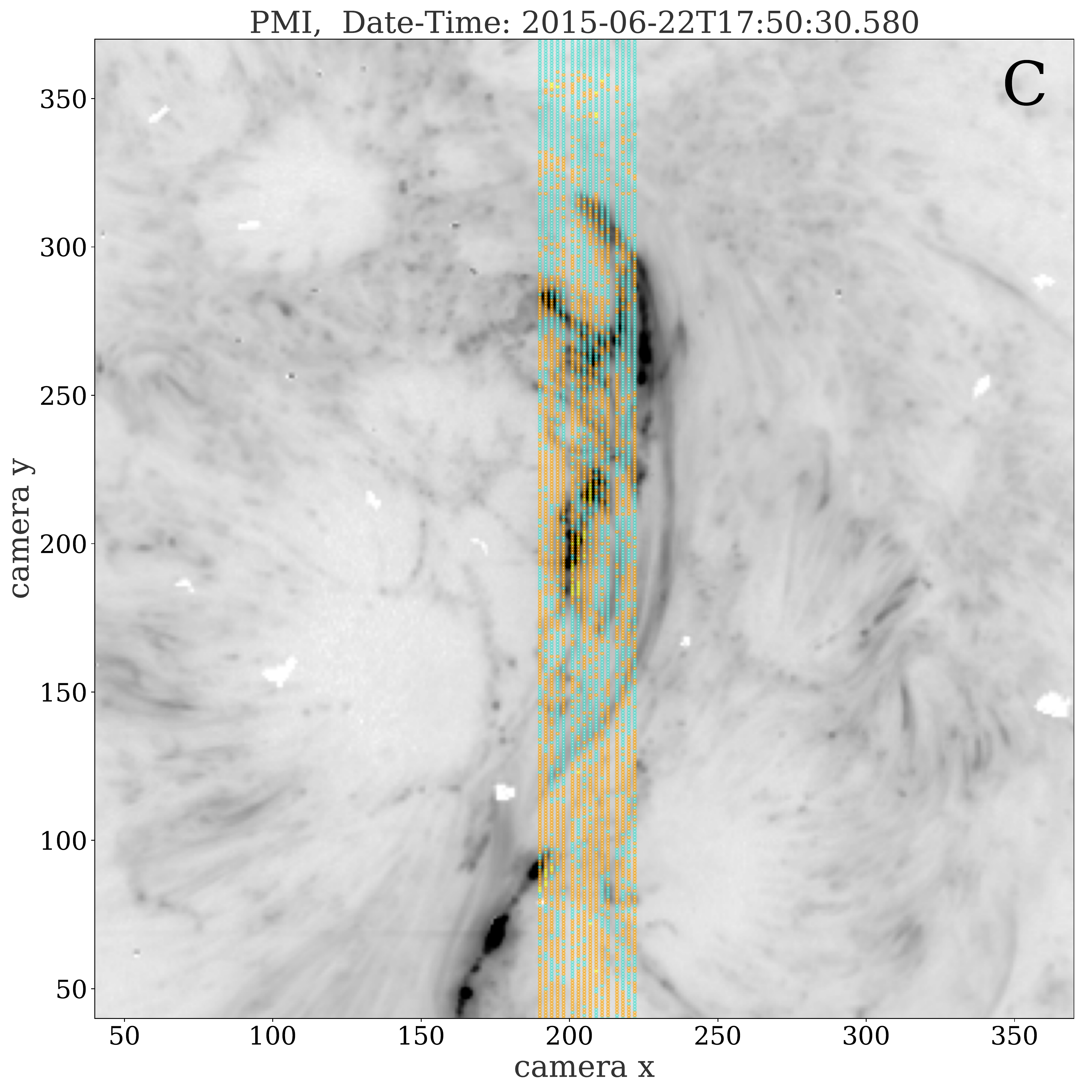}
\includegraphics[width=.33\textwidth]{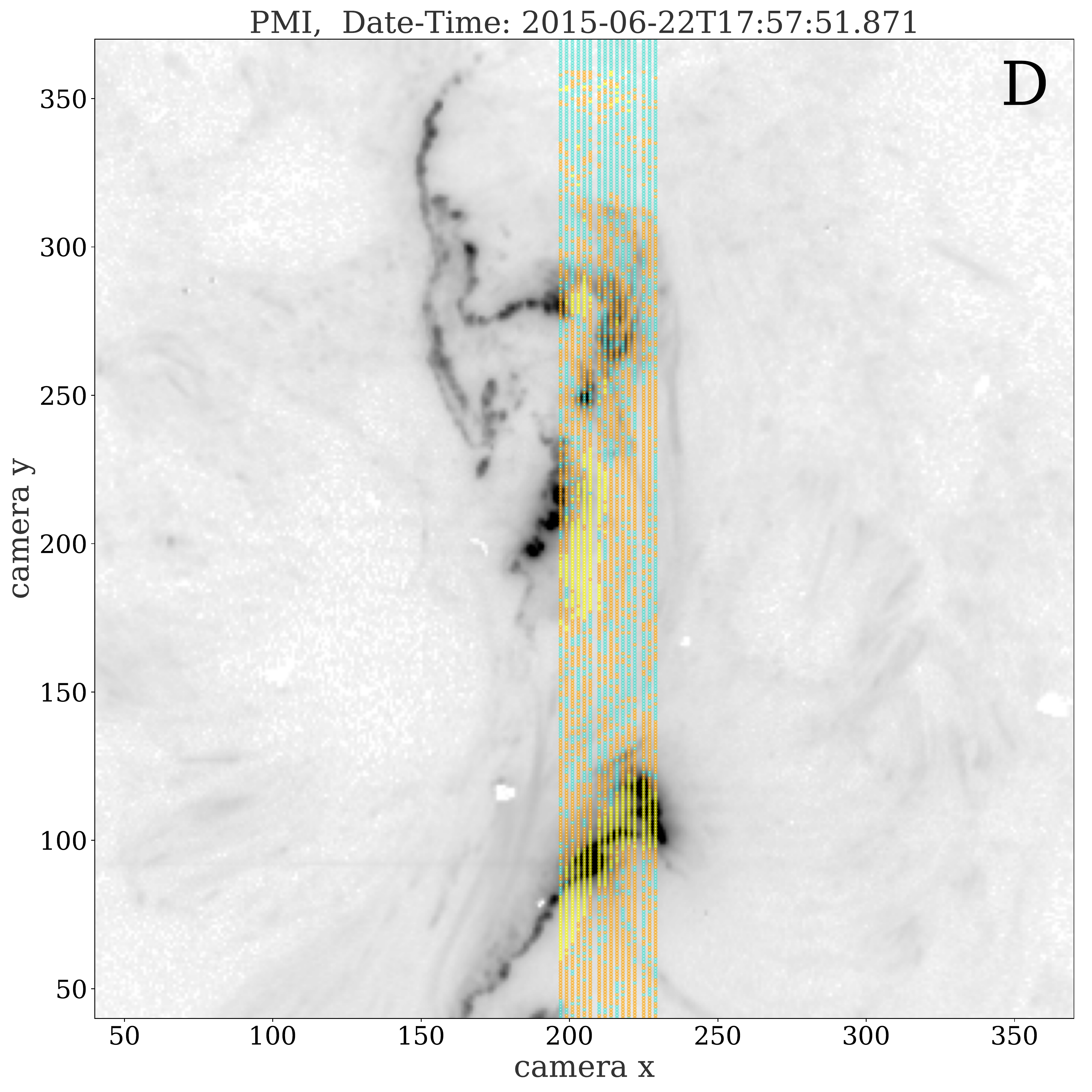}
\includegraphics[width=.33\textwidth]{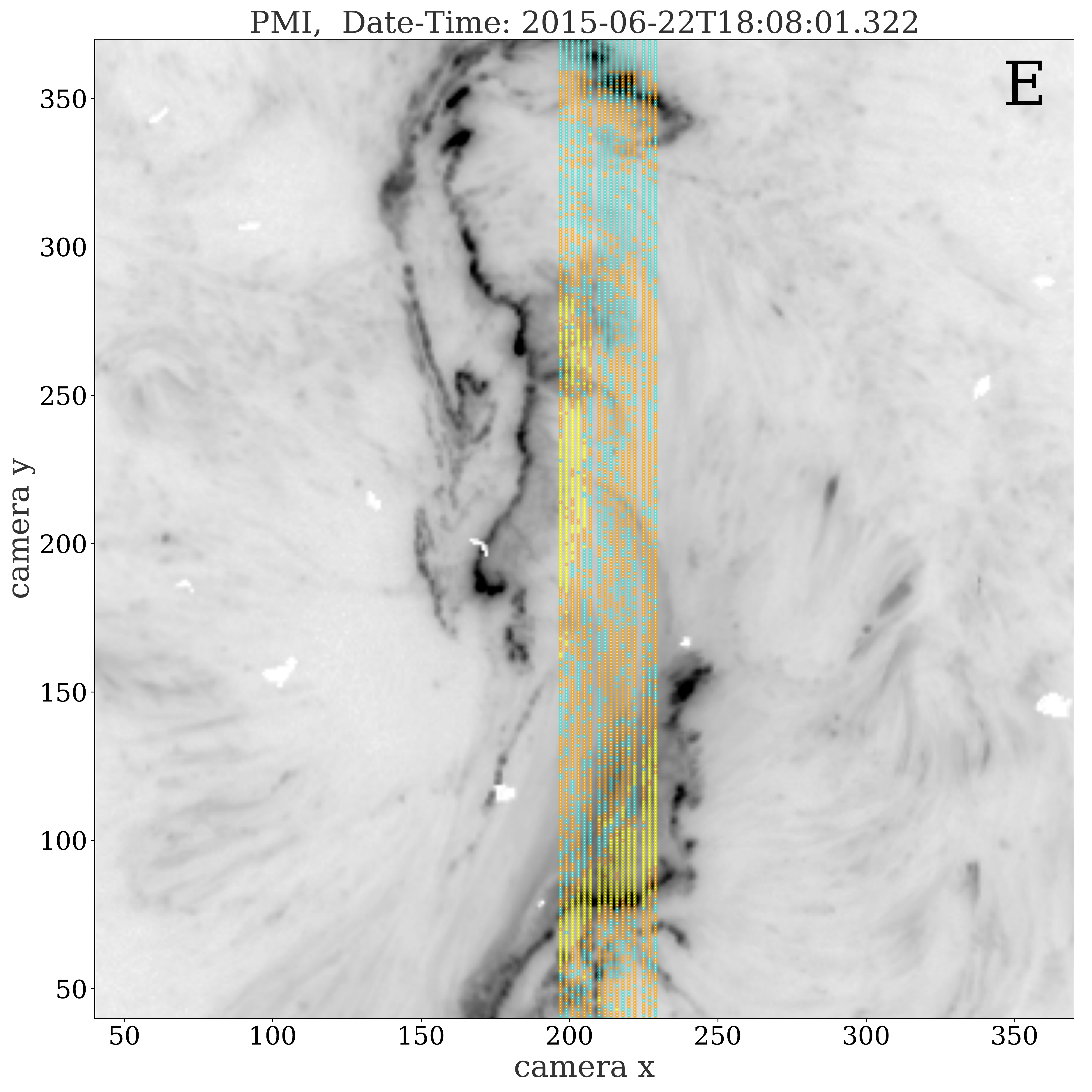}
\includegraphics[width=.33\textwidth]{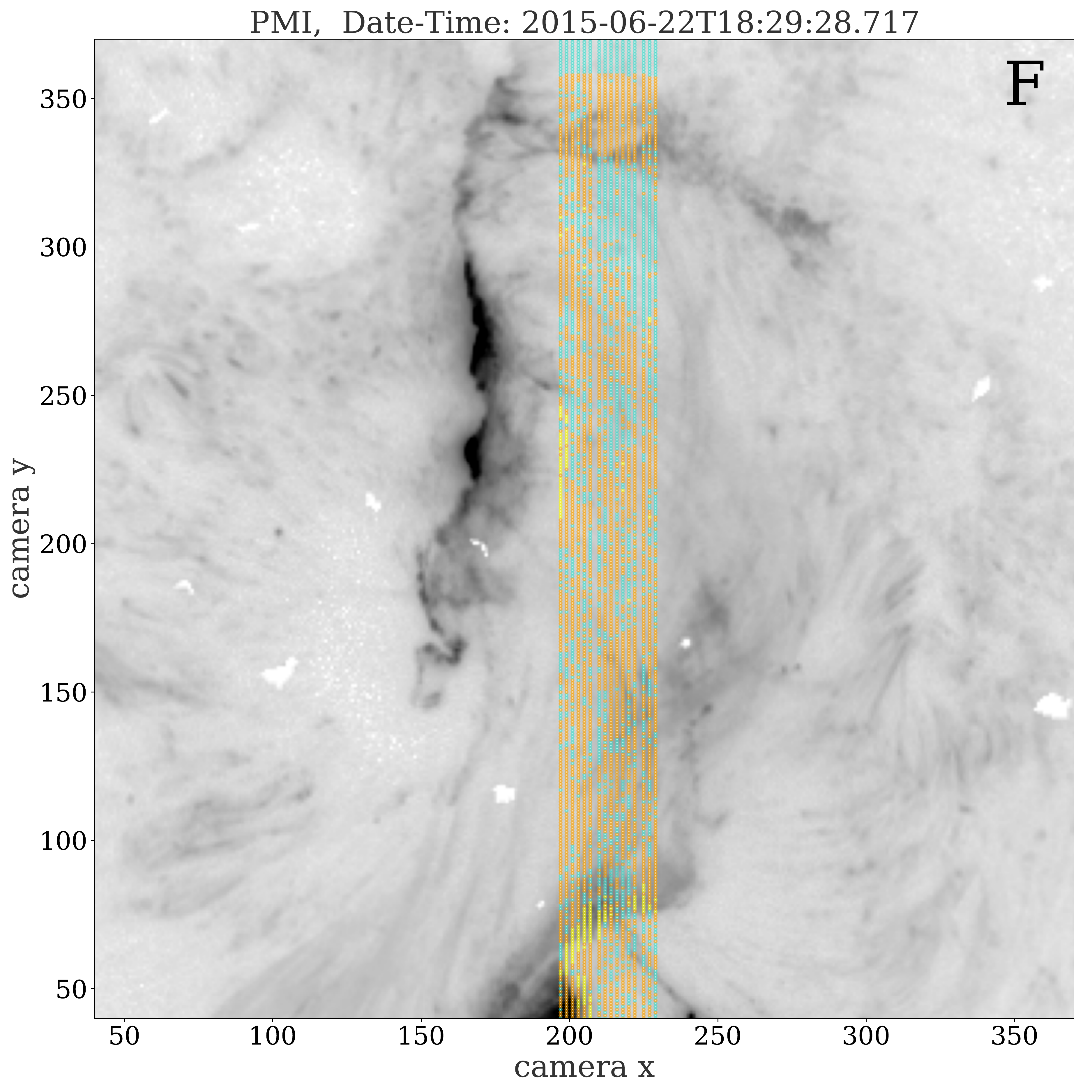}\\
\includegraphics[width=1\textwidth]{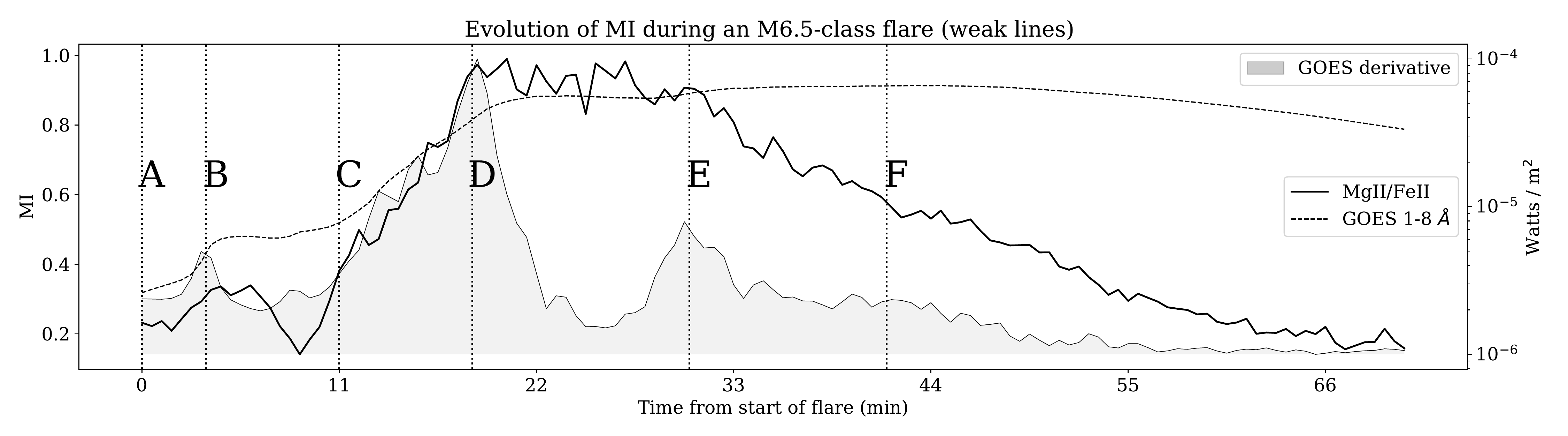}
\caption{Evolution of spectral correlations between \ion{Mg}{2} and \ion{Fe}{2} during an M6.5-class flare (\#13 in Table \ref{obs}). The upper 6 panels show the PMI associated with spectral pairs emanating from single raster sweeps. The color code splits the PMI into: Blue ($\text{PMI} \leq 0$), Orange ($0<\text{PMI}<1$) and Yellow ($\text{PMI}\geq 1$). The lower panel shows the GOES curve and derivative, with the accumulated MI scores of each SJI indicated by the vertical lines. We see that the MI is strongly correlated with the GOES derivative, and that the largest contributions to the MI are over the most energetic regions, i.e., the flare ribbons, meaning that \ion{Mg}{2} and \ion{Fe}{2} are correlated most strongly in flare ribbons.}
\label{PMI_SJI_EV}
\end{figure*}

To test this, we constructed an artificial data set consisting of 700 \ion{Mg}{2} and \ion{C}{2} spectra respectively, as seen in Figure \ref{pmi_synth}. Each line has two distinct types of spectral shapes labeled: Mg1, Mg2, C1 and C2. Note that there is a black profile in each of the four panels. This is referred to as the group centroid, and is taken to be the average of all the (orange or blue) profiles within the particular group. The colored bars on the bottom of the figure indicate whether the spectral shapes occur together or not. For instance, 500 spectra from the Mg1 group occur simultaneously with 400 spectra from group C1, similarly, 200 Mg2 spectra occur with 200 spectra from C2. This synthetic data set allows us to calculate the theoretical PMI of both (Mg1|C1) in blue, and (Mg2|C2) in orange, which can then be used as a benchmark. The convergence of the MINE-network for both cases can be seen in Figure \ref{convergence_of_pmi}. The theoretical PMI's derived from the artificial data set are indicated as horizontal lines color coded with the same scheme. The orange and blue curves represent the MINE-networks estimation of the PMI's between (Mg1|C1) and (Mg2|C2) respectively at each point of its 50 epoch training cycle. After 20 epochs, the network has converged on the theoretical values, with slight fluctuations persisting for the PMI between (Mg1|C1) in blue. We conclude that the PMI can safely be derived from the resulting hypothesis function of a carefully trained MINE-network.

\subsection{Single pixel correlations using PMI}
In Paper I, we postulated that the MI was concentrated over the flare ribbon. This argument was based on two observations: Firstly, that the maximum MI of all tested line-pairs coincided precisely with the peak of the GOES flux derivative, and secondly, that the MI of all line-pairs increased when sampling spectra only from the strict flare condition, rather than from the weak flare condition. The strict flare condition was designed to accept pixels that were related to \ion{Mg}{2} spectra which were drastically different from their quiet Sun counterparts. This difference was quantified using a variational autoencoder, see Paper I for details. It is still however an assumption that the spectra that differ most from quiet Sun profiles, in terms of shape, are confined exclusively to the flare ribbon. It might be the case that the spatial region which contributes the most to the MI is located just off of the ribbon. 

The ability to estimate the PMI between individual spectra allows us to directly test the hypothesis that the MI saturates directly over the flare ribbon. In Figure \ref{PMI_SJI_EV}, we calculated the PMI of each IRIS pixel in flare number 13 of Table \ref{obs} for the (\ion{Mg}{2}|\ion{Fe}{2}) line-pair. We then projected the results onto the corresponding slit jaw images (SJI) in terms of a color coding scheme, with blue indicating a $\text{PMI} \leq 0$, orange corresponding to $0<\text{PMI}<1$, and yellow with a $\text{PMI}\geq 1$. The lower panel indicates the GOES curve and derivative, with vertical lines labeled A through F specifying the times when the 6 different SJI images shown above were taken. This result indicates an explicit relationship between the lower and upper atmosphere that is induced either by large amounts of focused heating or bulk velocity flows. The spectral lines of \ion{Mg}{2} and \ion{Fe}{2} have maximum correlations directly over the flare ribbon, and during the impulsive phase, where energy deposition is at a maximum. This confirms that flare atmospheres are much more vertically coupled than quiet Sun atmospheres.

\section{Conditional probability}
\label{Conditional}
Including Paper I, so far we have calculated the total correlations (MI) between spectral shapes from different lines formed over different atmospheric heights, finding that some lines are more strongly coupled than other in flaring atmospheres. We then extended the methods to calculate the correlations on a spectrum-to-spectrum level (PMI). We discovered that most of the correlations appear directly over the flare ribbons. In what follows, we further develop these methods to determine the actual spectral shapes, so that we can address the question: given a \ion{Mg}{2} spectrum associated with the flare ribbons, what type of spectra occur co-spatially and co-temporally in \ion{C}{2}, \ion{Si}{4}, \ion{Fe}{2}, and \ion{Fe}{21}, and with what probabilities? To answer this question, we need to calculate the conditional probabilities (CPs) between pairs of spectra. In this section, we develop a simple method based on the k-means clustering algorithm, which will allow us to quickly derive the CPs with a minimal set of assumptions.
\begin{figure}[t] 
\centering
\includegraphics[width=.49\textwidth]{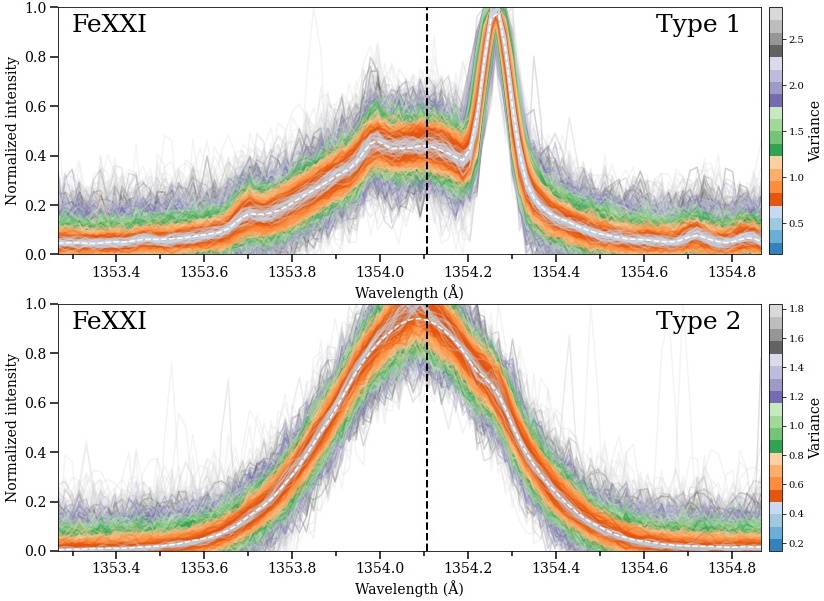}
\caption{An example of two different \ion{Fe}{21} spectral types. The upper panel shows a few thousand spectra and their variance away from the average spectral profile indicated as a white dashed line. The entire collection of spectra forms a spectral "type". In this particular case, the distinctive quality of this group is that the \ion{Fe}{21} and blended \ion{C}{1} emissions appear clearly distinct from one another. Similarly, the lower panel shows a different spectral type, where the \ion{C}{1} blend is entirely consumed by the \ion{Fe}{21} emission. Grouping spectra into similar shapes, regardless of their small variances, allows us to ignore irrelevant details, and perform statistical studies which were previously not possible.}
\label{spectral_types}
\end{figure}
\subsection{Deriving Conditional probabilities}
\label{CP_direct}
At high enough resolution, each spectrum is unique, therefore, it makes little sense to talk about a statistical study. In order to speak intuitively about a spectrum and its inter-line correlations, we have to introduce a degree of vagueness into its definition, so that when we talk about a spectral "type", we refer not to a precise spectrum, but rather to a collection of spectra with similar, but non-identical profile shapes, that nevertheless have some notable features in common. Ideally, all spectra that belong to a specific spectral type have a similar shape.

The k-means algorithm \citep{macqueen1967} is an ideal tool for automatically clustering spectra into groups of similar shapes. For a relevant and detailed explanation of the algorithm, see \cite{Panos_2018}. \Cref{spectral_types} shows two different \ion{Fe}{21} spectral types, with each group containing non-identical, but nevertheless similarly shaped profiles. Representing large amounts of slightly varying spectra as the same spectrum is an example of a vector quantization, and in the literature, the terms: type, group and label are often used interchangeably. A vector quantization can be thought of as a generalized form of data binning, where some information is sacrificed in order to gain purchase on the statistics of the problem. In this case, the small variations within each group are ignored.

\begin{figure}[t] 
\centering
\includegraphics[width=.49\textwidth]{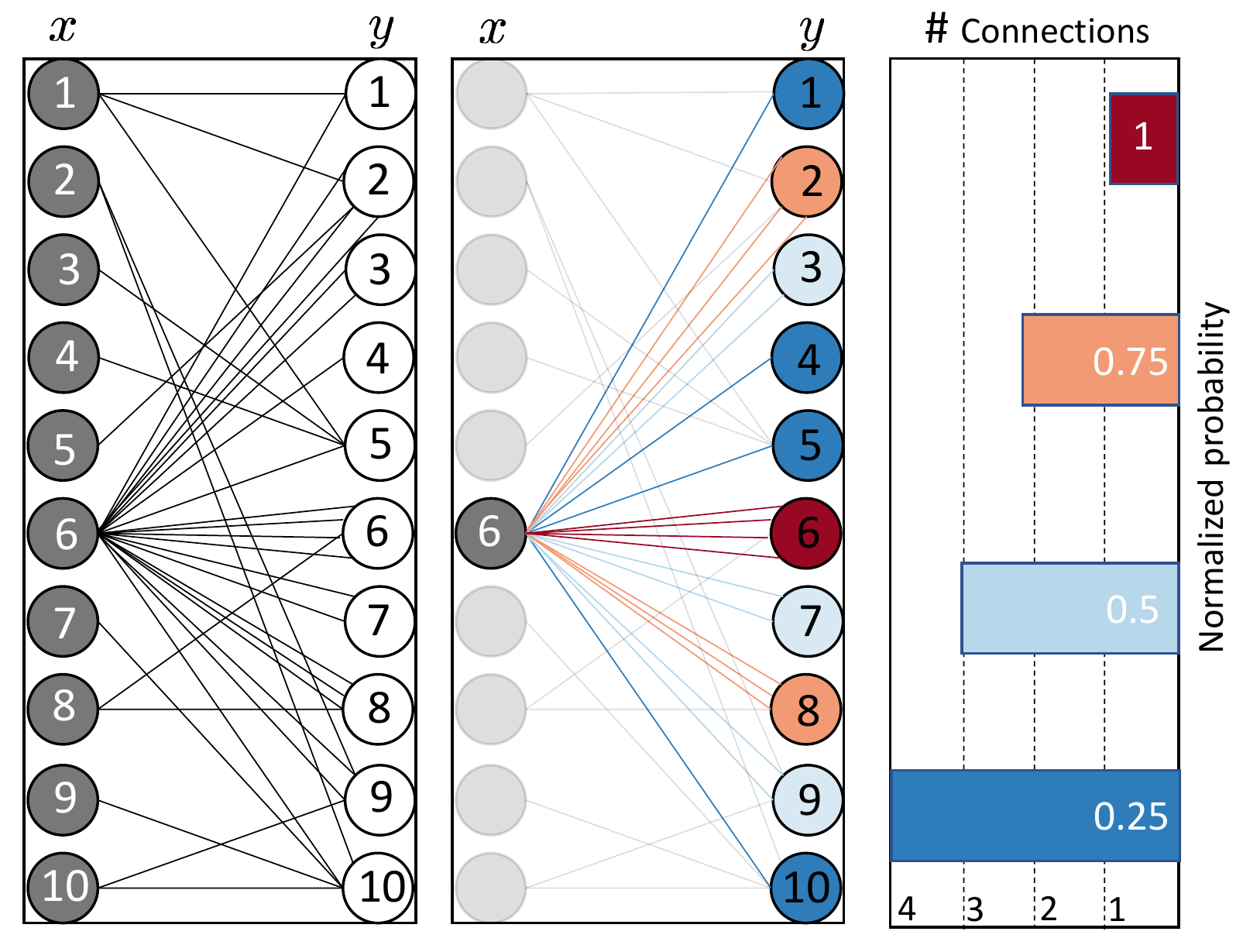}
\caption{Example of how to derive conditional probabilities to answer questions such as: given a certain \ion{Mg}{2} spectral type $x_i$, what is the probability of observing a specific type of \ion{C}{2} spectra $y_j$? The above figure shows 10 hypothetical spectral types for \ion{Mg}{2} and \ion{C}{2} represented in the left panel as grey and white circles respectively. Each spectral type contains thousands of similar but non-identical spectra. A particular CP-distribution can be calculated by counting the relative number of connections between a fixed grey node and the corresponding white nodes. The number of connections from one node to another represent the number of times spectra of type $x_i$ occur with spectra of type $y_j$ within the same IRIS pixel. In the middle panel, we have highlighted the connections for spectral type $x_6$, which may represent \ion{Mg}{2} spectra with large red wing enhancements. Most of the connections go to $y_6$, which most likely also contains \ion{C}{2} spectral shapes indicating downflows. The connections from $x_6$ to the other groups are weighted according to the total number of connections, allowing us to calculate the normalized CP-distributions seen in the right panel. The same color scheme is used when calculating the CP-distributions for a real fixed \ion{Mg}{2} spectral type in section \ref{Application_Section}.} 
\label{toy_condit}
\end{figure}

We used k-means to classify the spectra into 1000 spectral types for each of the 5 analyzed lines over the entire data set of 21 large flares. This resulted in $x_i, i\in{1000}$ groups for \ion{Mg}{2}, $y_j,j\in1000$ groups for \ion{C}{2} and so on. It is important to keep in mind that within any $x_i$ group, there are a large number of non-identical, but similar shaped spectra. This inter-group-variance implies that the derived CPs depend strongly on the number of spectral types, however, this variance becomes insignificant when selecting a large number of groups. The CP-distributions can then be derived via a simple counting scheme demonstrated in Figure~\ref{toy_condit}.

This figure shows  a  hypothetical example that uses 10 spectral types for both $x$ and $y$ instead of  the 1000 groups used on the actual data. In this example, $x$ and y could refer to \ion{Mg}{2} and \ion{C}{2} spectra respectively. The left panel of Figure~\ref{toy_condit} is a diagrammatic realization of how many times a spectrum from one \ion{Mg}{2} group appears simultaneously with spectra from a \ion{C}{2} group. For example, spectra from group $x_1$ occur once within the same IRIS pixel as spectra from groups $y_1, y_2$ and $y_5$. Therefore, the conditional probabilities are $p(y_1|x_1)=p(y_2|x_1)=p(y_5|x_1)=0.33$, with all other conditional probabilities being zero. The middle panel highlights the connections for the more complicated case of fixed $x_6$. The \ion{Mg}{2} spectra in this group appear to be connected to a variety of different \ion{C}{2} groups with different frequencies. For instance, $x_6$ occurs with $y_6$ four times while only once with spectra from group $y_1$. The frequencies of connections (CPs) are displayed on the vertical axis in the panel on the right, with the highest conditional probability indicated in red as $p(y_6|x_6)=0.2$ and only occurring once. Note that we have chosen to normalize the CPs by the maximum value $0.2$, derived from 4 occurrences of a total 20 connections. The lowest conditional probability on the other hand is $0.05$, and is realized four times by single connections between $x_6$ and $y_1, y_4, y_5$ and $y_{10}$, leading to the normalized probability of 0.25 in the figure. Note that for this particular case, the connections in the middle panel as well as the probability density of the right panel is depicted in dark blue. We will adopt an identical normalization and color scheme when calculating the CP-distributions for real data.

On a technical side note, the high number of spectral types means that some distinct groups share similar looking spectra. In order to properly estimate the CPs, these groups were manually merged, leading to fewer than 1000 \ion{Mg}{2} spectral types in the end.

\begin{figure*}[t] 
\centering
\includegraphics[trim={5cm 2cm 8.8cm 3.5cm},clip,width=.49\textwidth]{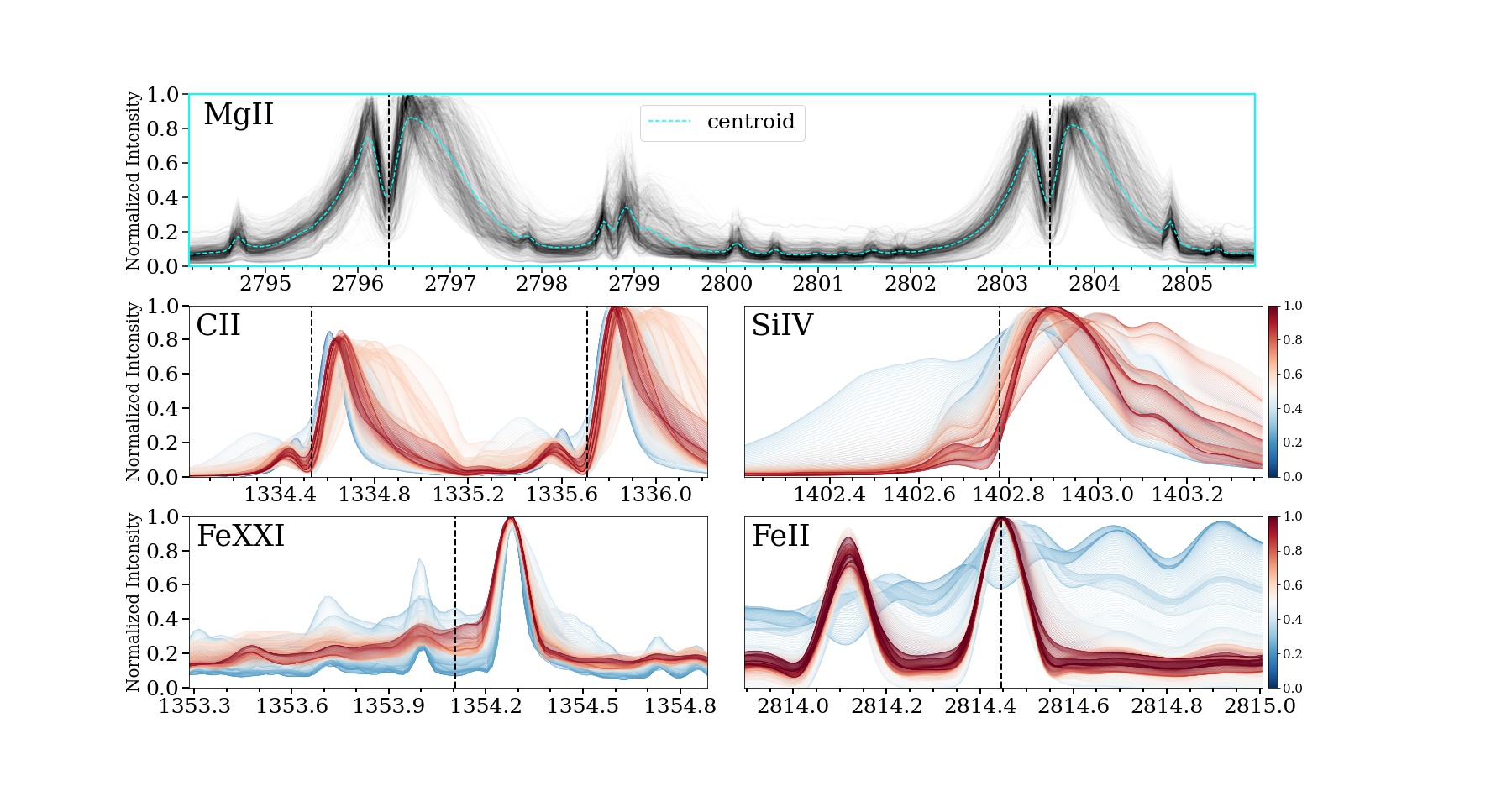}
\includegraphics[trim={5cm 2cm 8.8cm 3.5cm},clip,width=.49\textwidth]{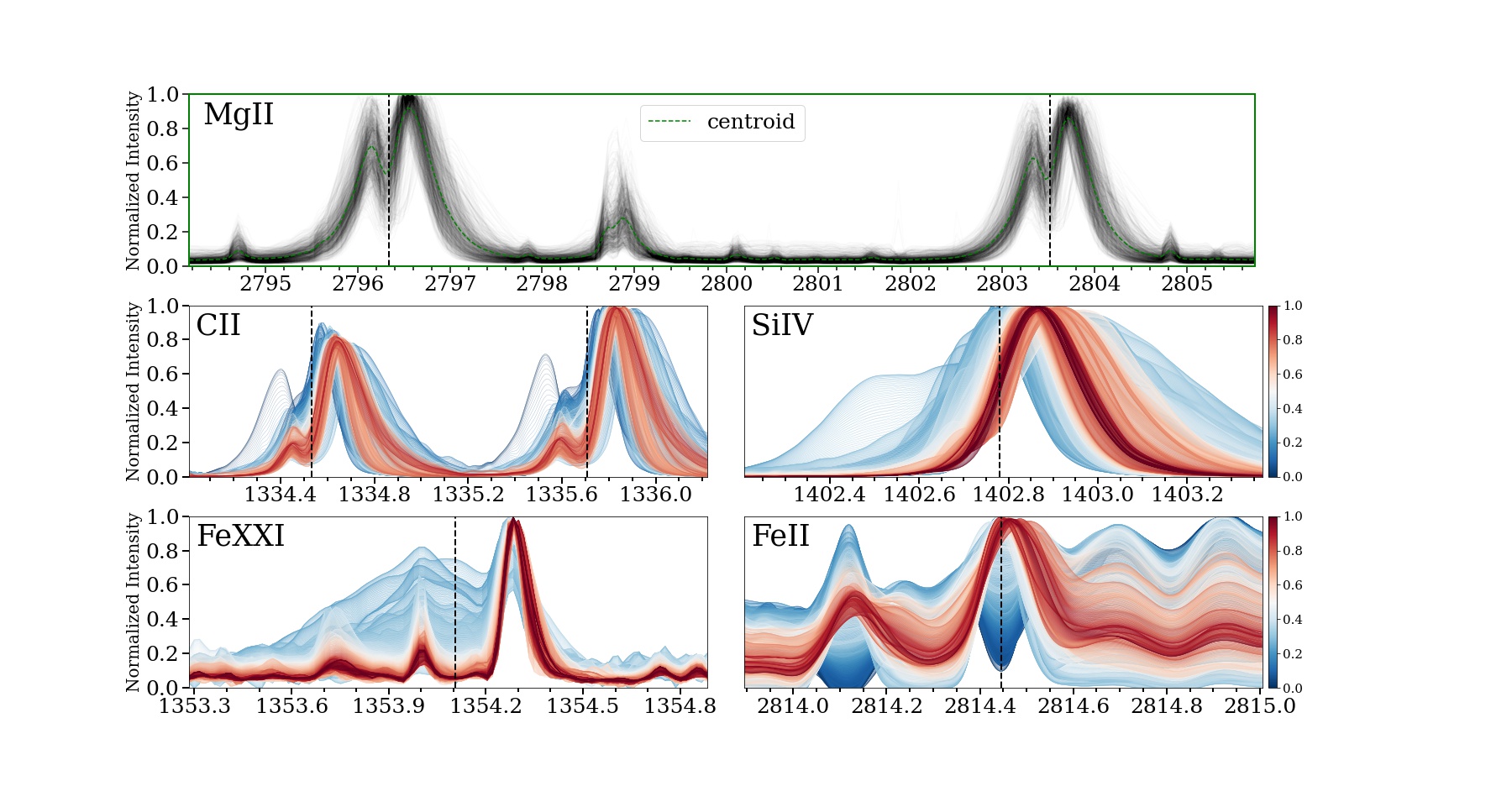}
\includegraphics[trim={6cm 2cm 9cm 3.5cm},clip,width=.49\textwidth]{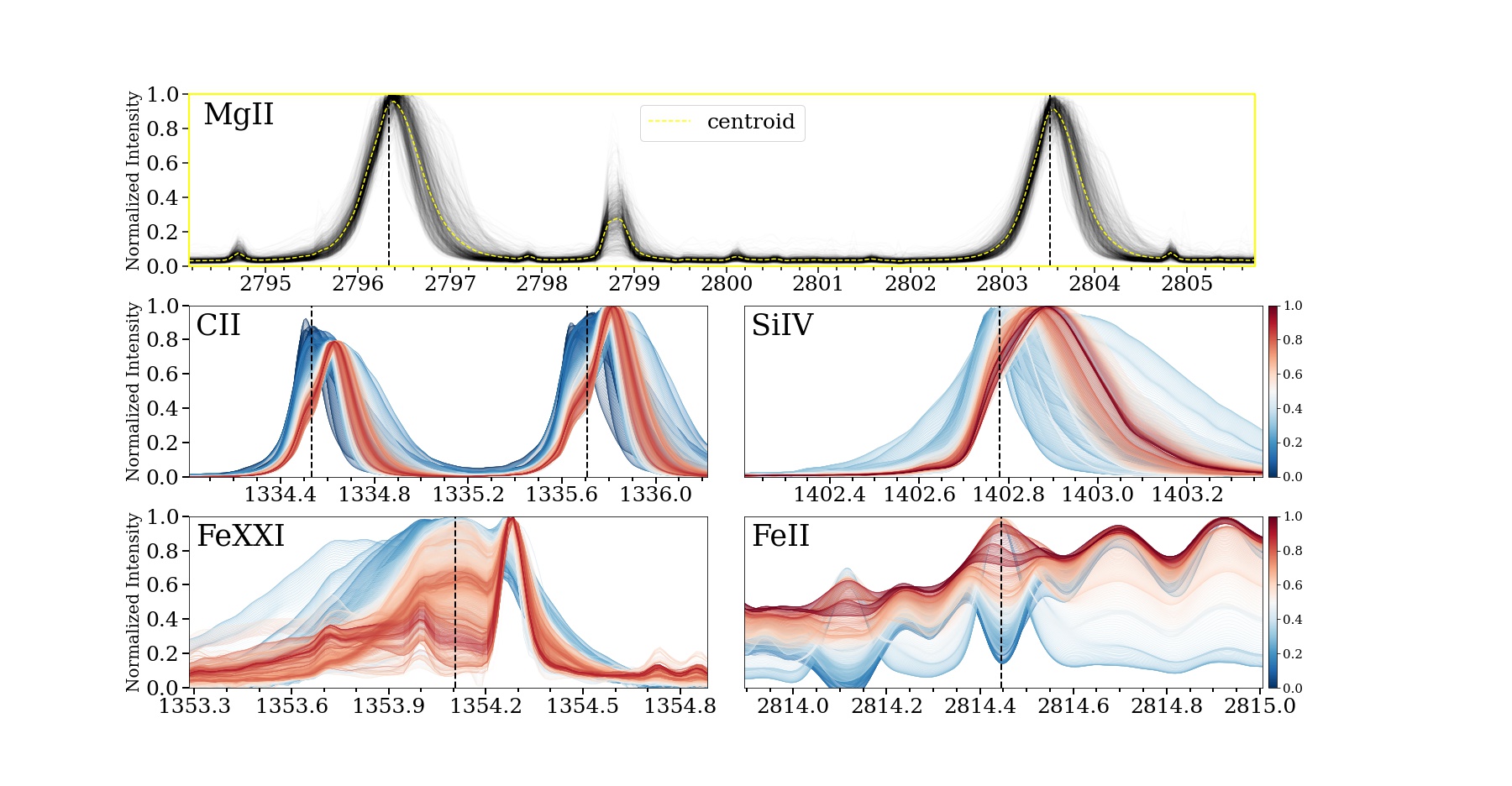}
\includegraphics[trim={6cm 2cm 9cm 3.5cm},clip,width=.49\textwidth]{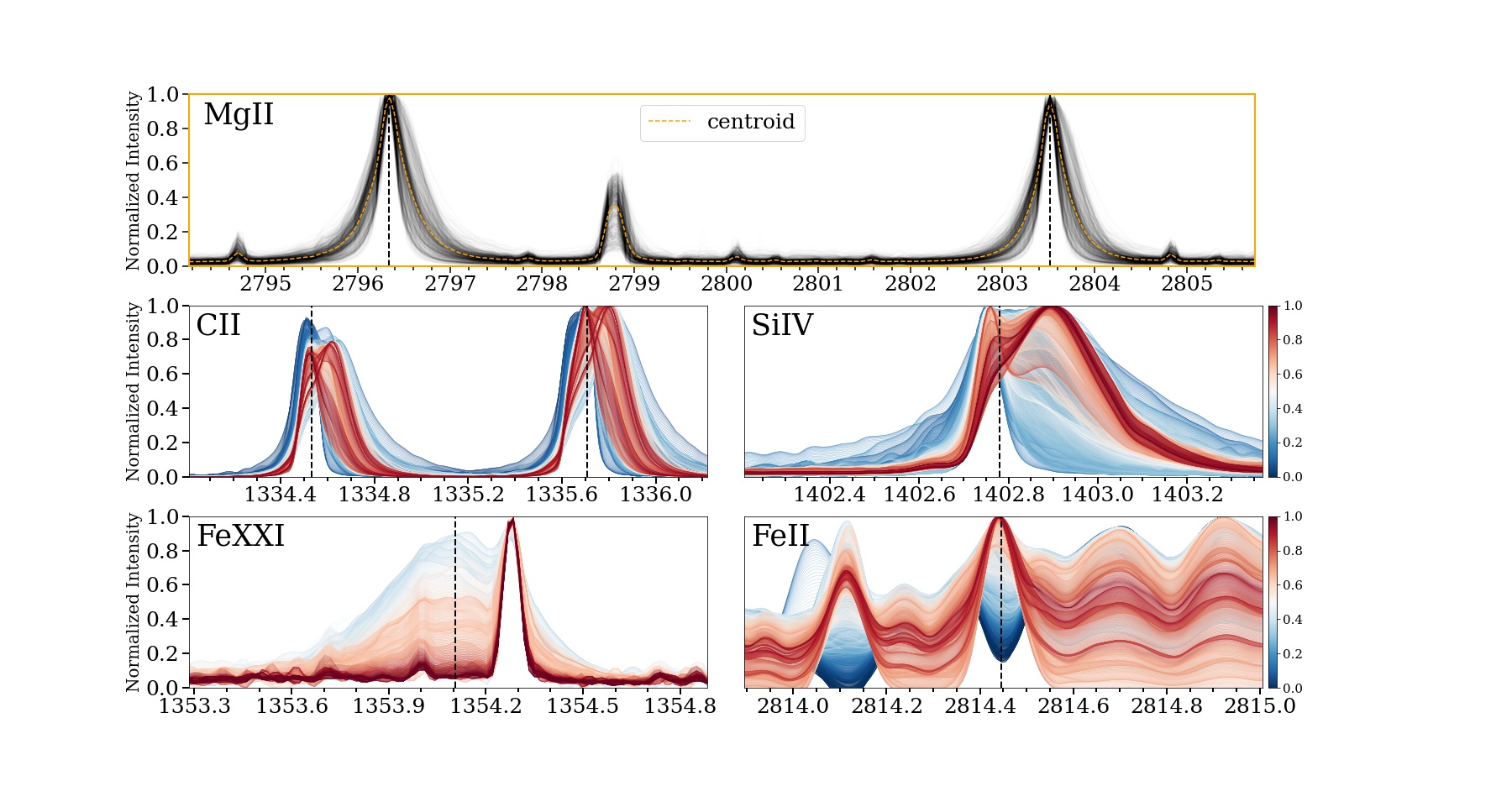}
\includegraphics[trim={6cm 2cm 9cm 3.5cm},clip,width=.49\textwidth]{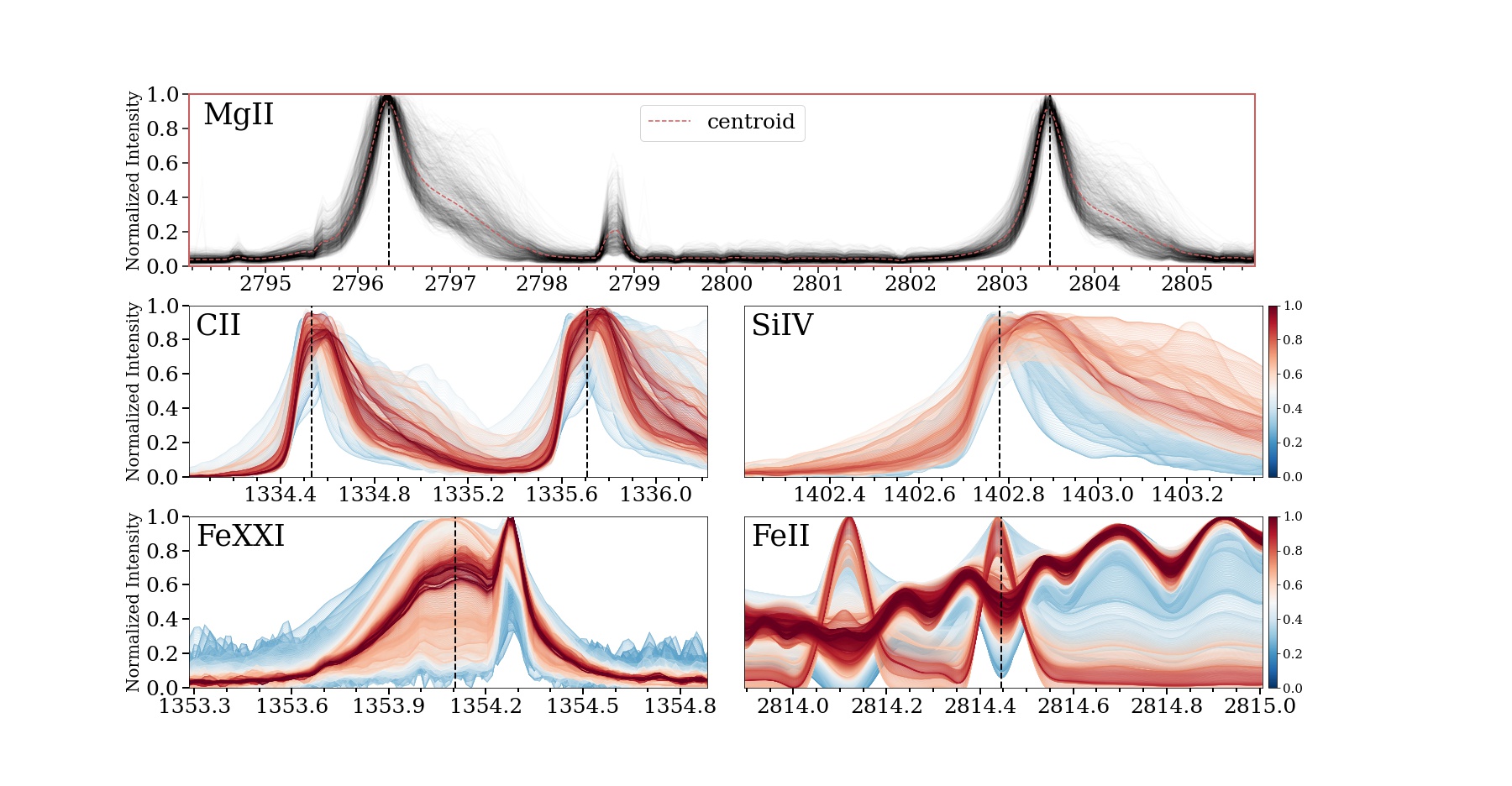}
\includegraphics[trim={6cm 2cm 9cm 3.5cm},clip,width=.49\textwidth]{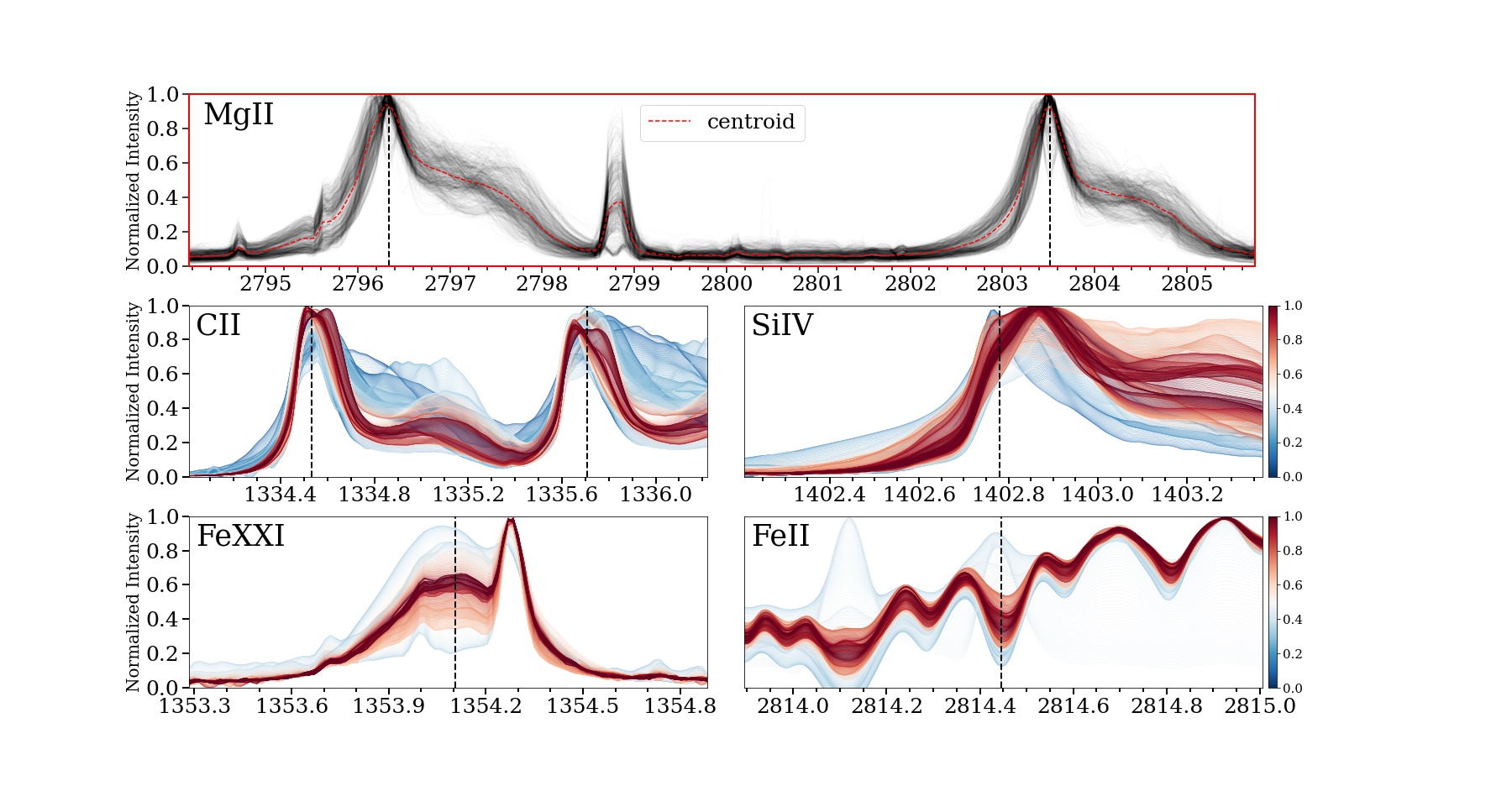}
\caption{Each of the 6 color-coded panels shows a particular \ion{Mg}{2} spectral type and their most statistically likely responses over the remaining 4 spectral lines within the sub panels. We have provided the following naming convention for each category: \textit{Strong-Ribbon-Front} (upper left) and \textit{Weak-Ribbon-Front} (upper right), \textit{Broad-Single-Peaked} (middle left) and \textit{Narrow-Single-Peaked} (middle right), \textit{Downflow} (bottom left) and \textit{Downflow-Extended-Shoulder} (bottom right). The line cores are indicated by vertical dashed lines, and the color bars represent the normalized conditional probabilities of each profile, with redder profiles being more statistically likely than bluer profiles. From the upper left panel, it is clear that reversals with a peak asymmetry occur simultaneously in \ion{Mg}{2}, \ion{C}{2} and \ion{Si}{4}. Additionally, \ion{Fe}{2} is very likely to be in emission during the impulsive phase, indicating strong heating of the lower atmosphere. \ion{Fe}{21} appears with a delay, going from blueshifted to stationary as can be seen from the top-left to bottom-right.} 
\label{probA}
\end{figure*}

\begin{figure*}[t]
\centering
\includegraphics[trim={0cm 0cm 0cm 0cm},clip,width=1\textwidth]{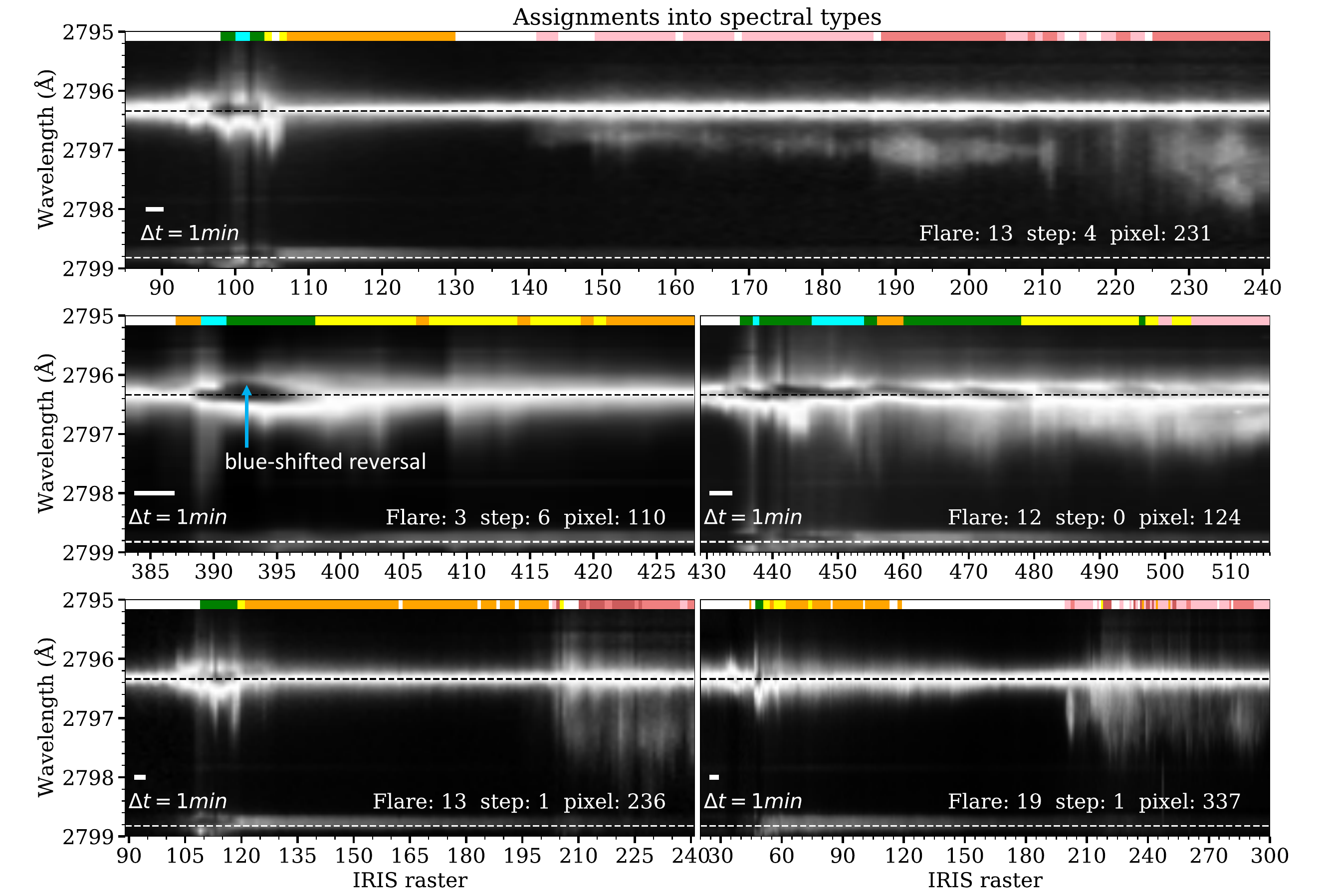}
\caption{Temporal occurrence of the different flare spectra (single-pixel). Each panel shows a different \ion{Mg}{2} spectrogram covering both the k-core and red wing subordinate line, with the rest wavelengths indicated by black and white horizontal dashed lines respectively, and each time step normalized separately to one. The flare number, raster step position, and pixel are indicated in the lower right hand side of each panel. A white stripe in the lower left coroner shows the length of a minute interval. The colored squares at the top of each plot assign the spectra directly below them into one of the 6 spectral types in \Cref{probA}. Spectra belonging to white groups were not analyzed here. The cyan and green spectral types consistently flag the impulsive phase at the beginning of each flare sequence, and consist of a blue shifted central reversal indicated in the first middle panel by a blue arrow, while downflows (red colors) appear at later stages.}
\label{Temporal_evolution}
\end{figure*}

\begin{figure*}[t] 
\centering
\includegraphics[width=.49\textwidth]{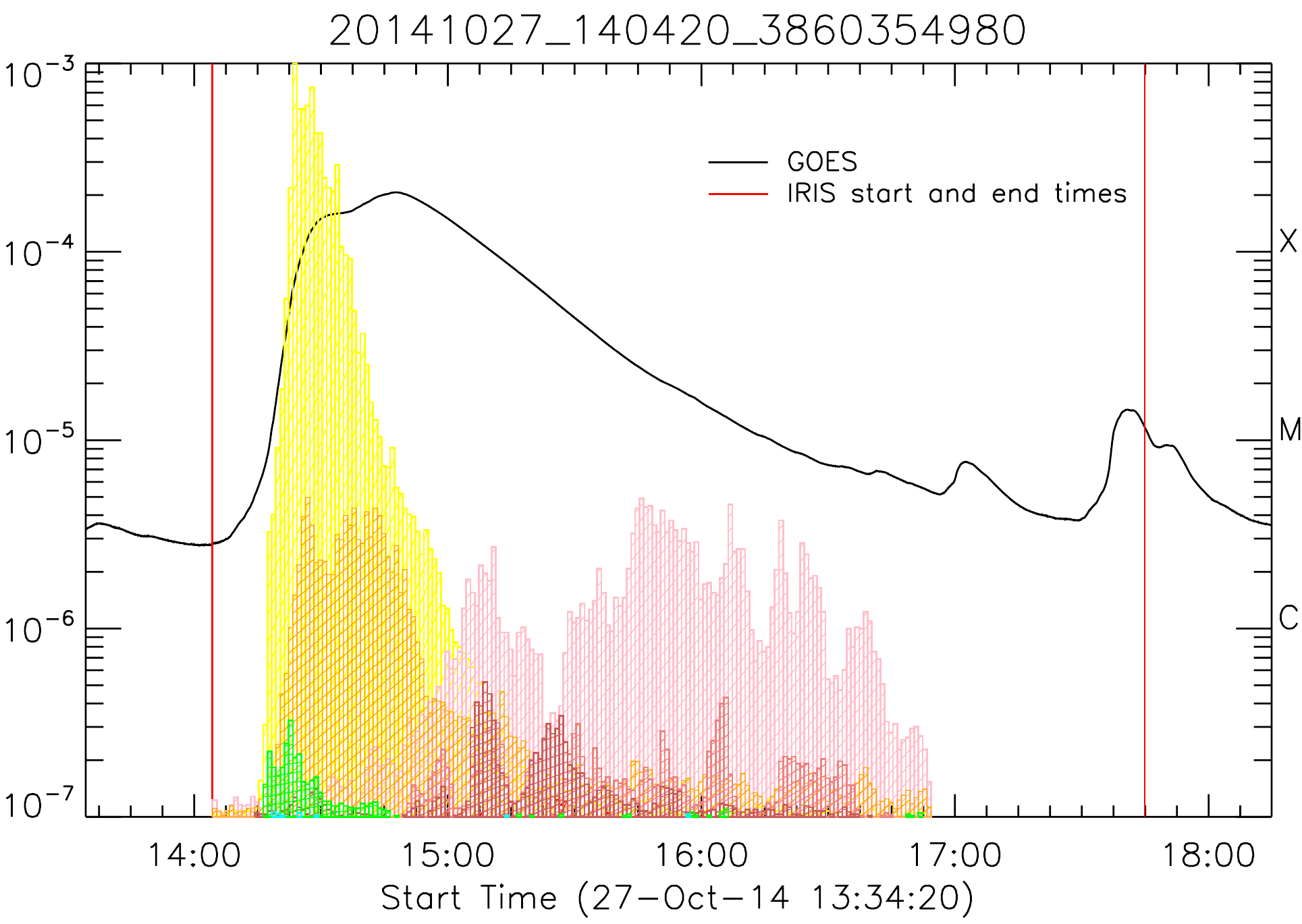}
\includegraphics[width=.49\textwidth]{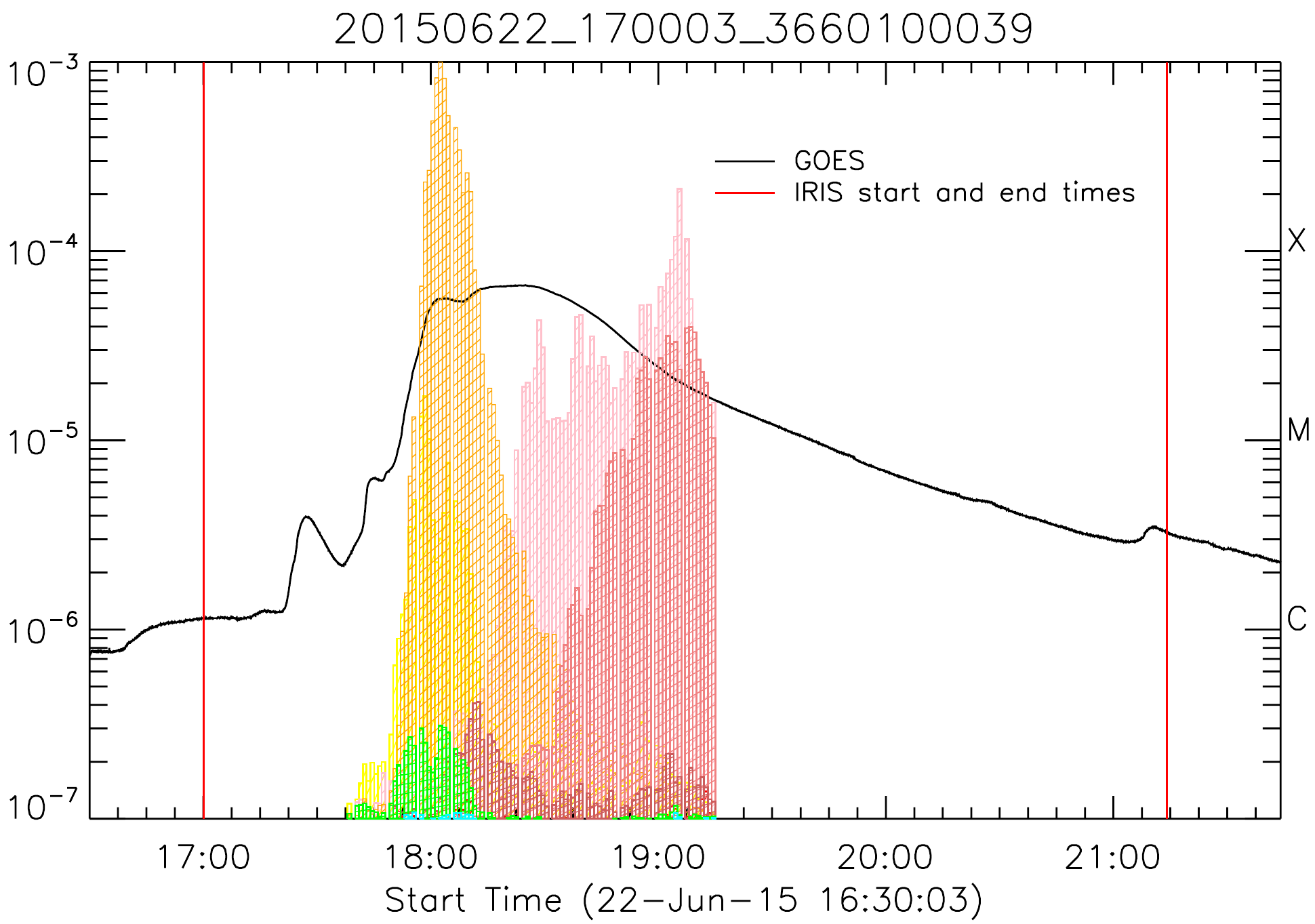}
\caption{Temporal occurrence of the different flare spectra (all-pixels). The histograms depict color-coded types of spectra, from ribbon front (cyan, green), to single-peaked (yellow, orange) and downflows (red shades). The GOES X-ray flux is shown in the background and the red vertical lines indicate the start and end times of the IRIS observation, which does not necessarily coincide with the manually selected analysis range. The sequence seen in \Cref{Temporal_evolution} appear to be true for the majority of flare related pixels. The larger overlap between spectral types is due to the gradual unfolding of the flare ribbons, something that a single pixel analysis is not affected by. It is evident that all spectral types analyzed here are correlated to flares, and that downflows appear during later flare phases.}
\label{temporalevol}
\end{figure*}

\section{Application to \ion{Mg}{2} Spectra}
\label{Application_Section}
In this section, we apply the above method to derive the CPs of 6 archetypal \ion{Mg}{2} flare spectra, and analyze the responses in the remaining spectral windows. The results can be seen in \Cref{probA}, where each of the six panels correspond to a different \ion{Mg}{2} spectral type: \textit{Ribbon-Front-Spectra} (cyan and green), \textit{Single-Peaked-Spectra} (yellow and orange), \textit{Downflows} (pink and red). The most likely spectral responses for each \ion{Mg}{2} type can be seen within the sub-panels. Redder spectra are more likely than spectra in blue shades. As an interpretive example, if we observe a \ion{Mg}{2} spectrum that looks similar to the profiles shown in the upper cyan panel, then it is almost guaranteed that \ion{C}{2} has a large peak asymmetry, and that \ion{Fe}{2} is most likely in emission.

A single panel typically contains thousands of spectra, we therefore created a compression scheme that merged similar spectra together for the figure while still retaining any important variances from less probable profiles. The CPs were updated according to the average score of the merged spectra.

\subsection{Temporal behavior of the spectra}
In addition to analyzing the CP-distributions of \Cref{probA}, analyzing the temporal ordering of the spectra in relation to the impulsive phase, allows us to interpret the spectral types within the context of the evolution of a flare. We therefore marked the locations of each spectral type on Mg II spectrograms taken from all of our 21 flares.

A few examples that cover both the \ion{Mg}{2} k-core and red wing subordinate line can be seen in \Cref{Temporal_evolution}, with the rest wavelengths indicated by dashed black and white horizontal lines, and the flare number, raster step (with the count starting at 0), and pixel along the slit denoted in the lower right hand corner of each panel. The colored squares located above each spectrum indicate which spectral type it belongs to. For instance, the spectra in the upper panel directly below the cyan and green squares at about the 100th raster mark belong to the \textit{Ribbon-Front} type spectra of \Cref{probA}. Because the observations raster step is an ambiguous measure of time, we have included a reference for the length of a minute represented in the lower left hand corner of each panel by a solid white stripe.

\Cref{Temporal_evolution} shows that there is a weak mixing in the time domain, so that the spectral types are not consistently ordered in relation to the impulsive burst picked up by the cyan and green spectra. This can be seen in both the middle left and lower right spectrogram, where orange types appear briefly as the first spectra in the sequence. Additionally, orange and yellow types can flicker between one another, while downflows seem to be confined to the end of the flare sequence, but need not be visible in every case. This mixing must be taken into account when interpreting the CP-distributions of \Cref{probA}.

Despite the weak mixing, there are some general characteristics that can be extracted. As the flare ribbons deposit their energy, the \ion{Mg}{2} spectra develop central reversals which often appear blue shifted. This blue shift is most notable in the two central panels of \Cref{Temporal_evolution}, where the dark section in the middle of the spectrogram (reversal) is biased towards smaller wavelengths (see blue arrow). Flare 12 was confined between two sunspots, which may explain the oscillatory behavior of the reversal, since each strand would deposit pulses of non-thermal electrons as the reconnection site propagates upwards. During the time interval of the burst, we see that the triplet line also goes into emission on account of a shared formation height with the core during the early flare heating phases. Most panels also show a strong peak asymmetry for the cyan and green profiles, with an enhanced $2\text{k}\text{r}$ peak. After the initial shock, the \textit{Ribbon-Front-Spectra} relax back to line center and become \textit{Single-Peaked} (yellow and orange) before developing extensive red wing emissions from coronal rain. Other authors have noted similar line behavior over hard X-ray footpoints \citep{Liu_2015}. {Importantly, the \textit{Ribbon-Front-Spectra} typically have a lifetime of 1-3 minutes and can evolve from single peaked spectra. Therefore, the reversal is not a consequence of a typical quiet Sun profile being recoupled to the Planck function, but rather the result of dynamical motions. More studies should be undertaken to accurately measure the relaxation time of the central reversal.}

This typical sequential behavior can be seen at larger scales in \Cref{temporalevol}, where we show the normalized temporal occurrence of each spectral type for two flares, taken not just with respect to a single pixel, but aggregated over all pixels. The black line denotes the GOES X-ray flux (1-8 \AA) and the red vertical lines indicate the start and end times of the IRIS observations, of which we only analyzed a duration around the flare. Again, we see that the \textit{Ribbon-Front} spectra (cyan, green) are located mostly within the impulsive phase, followed by single peaked spectra (yellow, orange) and then during the gradual phase we see many downflow counts. Because the ribbon unfolds over time, each sequence is initiated with a slight delay time to the previous sequence, explaining why there is more overlap between the spectral types than in \Cref{Temporal_evolution}.

 Although the majority of \textit{Ribbon-Front-Spectra} occur in the impulsive phase, there are many instances, which can be seen here, where similar spectral shapes appear during the gradual phase, often in locations corresponding to clear downflows within the 1400 SJI of IRIS. Although these late phase spectra are grouped together with the \textit{Ribbon-Front} spectra, they often appear with much lower intensities and smaller triplet emission.

{The results from the above sequence plots urge cation when interpreting the CP-distributions of \Cref{probA}, since we might be conflating line responses from different parts of the flare sequence within the same panel.}

\subsection{Ribbon-Front-Spectra}

The \textit{Ribbon-Front-Spectra} of \Cref{probA} come in two flavors: A \textit{Strong} variant in the upper left cyan panel, and a \textit{Weak} variant in the upper right green panel. These spectral types were found at the moving edges of flare ribbons \citep{Xu_2016,rubiodacostaetal2016}, and are now recognized as a common feature during the impulsive phase of many flares \citep{Panos_2018}. This is now doubly confirmed by \Cref{Temporal_evolution}. They have very broad non-thermal widths, asymmetric peaks, and slightly blue shifted central reversals, whose formation properties are still under debate. 

\subsubsection{Formation properties of the blue-shifted-reversal}

The blue-shifted reversals may result from a superposition of unresolved downflows at different heights \citep{Rubio_da_Costa_2017}, but simple non-LTE cloud models have also reproduced the observed peak asymmetries with upflows \citep{Mg_ran1}. The simplicity of these cloud models unfortunately do not allow us to determine the likelihood for the upflow scenario. Simulations have demonstrated a possible third mechanism based on a PRD effect \citep{Kerr_mgII_1}. In this third model, the blue and red peak, which outline the central reversal arise from two different processes. The enhanced $2\text{r}$ peaks may be constructed out of red Doppler-shifted photons due to a dense downward propagating condensate triggered by the reconnection event. This condensate would naturally displace the maximum opacity into the red, allowing the bluer photons that constitute the $2\text{v}$ peaks to have longer mean free paths. In combination with a red shifted absorption profile, coherent scattering from PRD may allow the blue photons to remain in their optically thinner tracks, without being reabsorbed and emitted at redder frequencies. This interplay between a red shifted absorption profile and coherent PRD scattering, may thin the atmosphere for bluer wavelengths, and free enough photons to form the observed $2\text{v}$ peak. It appears that PRD, at least for the \ion{Mg}{2} lines, could be important for the blue peak formation, which is observed to be a combination of both optically thick and thin line emission \citep{Kerr_mgII_1}.

Our results in \Cref{probA} show that these types of spectral profiles are not unique to \ion{Mg}{2}, but also simultaneously appear in both \ion{C}{2} and \ion{Si}{4}. PRD effects are not considered to be important for both of these lines within the current modeling literature \citep[e.g.,][]{CII_qs1, Thick_SiIV}. It may therefore be the case that under the extreme circumstances of a flaring atmosphere, the correct modeling of these lines also requires the inclusion of PRD effects, however, we find this difficult to believe, since if anything, the density rich flaring atmosphere should increase the photon destruction probability $\epsilon=C/A_{ij}$, leading to a reduction of coherence in the laboratory frame. Alternatively, the larger mean free paths $\tau$ of \ion{Si}{4} may not require the additional action of coherent scattering via PRD to unlock and free blue photons, and a simple displacement of the absorption profile from line center might suffice. Considering that we find very similar spectral shapes with blue-shifted reversals during later phases of flares where plasma is seen to fall towards the solar surface, downflows at least seem to be a more likely explanation of the reversals than upflows.

\begin{figure}[t] 
\centering
\includegraphics[trim={.5cm 0cm 0cm 0cm},clip,width=.5\textwidth]{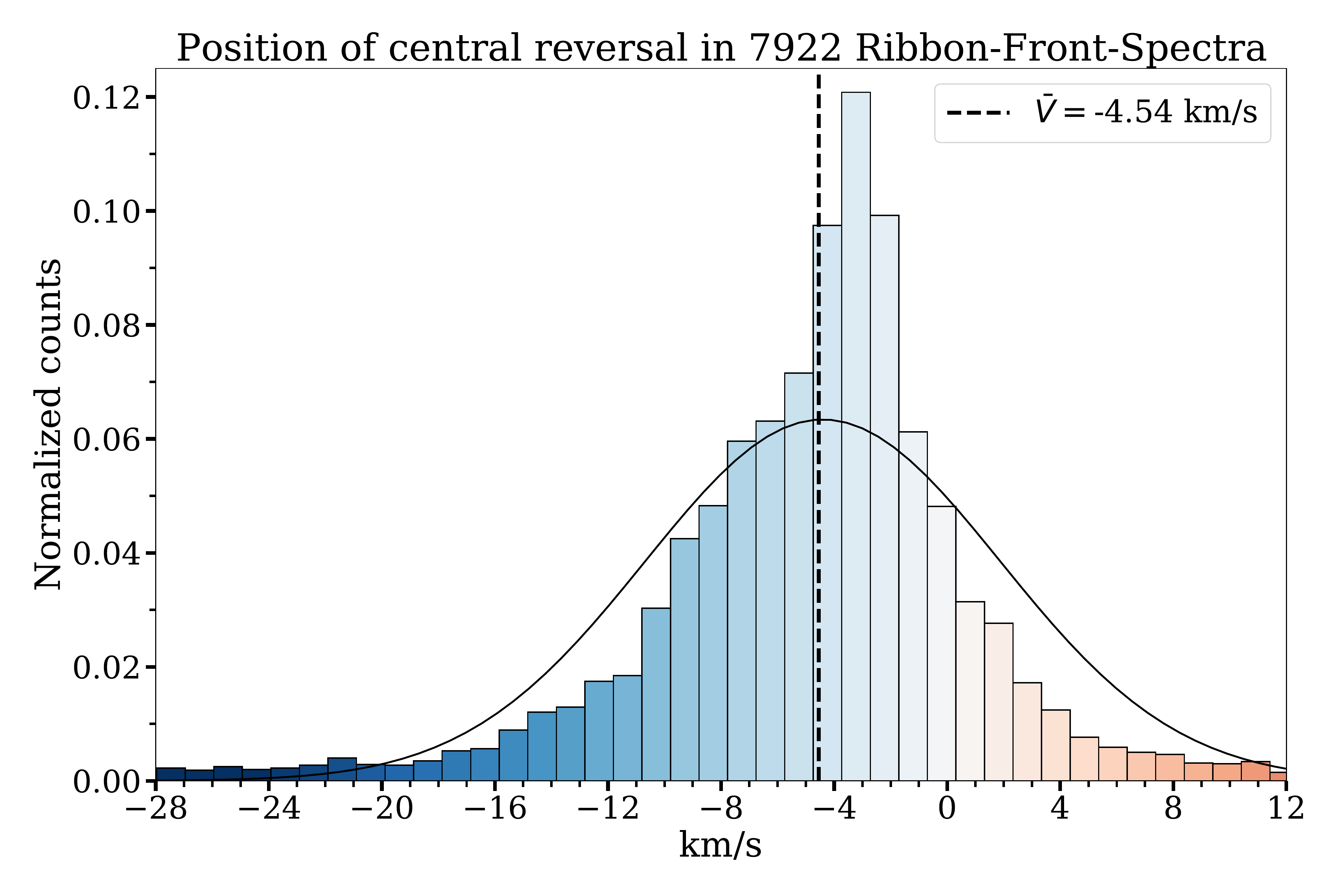}
\caption{Velocity distribution of the central reversals for all ribbon front spectra corresponding to the cyan and green groups. The minima of the reversals were located below the resolution limit by interpolating a small window around the core and fitting the reversal with a higher order polynomial. Negative values represent blue wavelength shifts, while positive value  correspond to red shifts. We find that the majority of \textit{Ribbon-Front} spectra have a central minimum that is blue shifted with a large spread.}
\label{Central_Dist}
\end{figure}

\subsubsection{Velocity statistics of the reversal}

The distribution relating to the position of the blue shifted reversal for the collection of all cyan and green \textit{Ribbon-Front-Spectra} can be seen in \Cref{Central_Dist}. The locations of the reversals were calculated below the resolution limit by interpolating the profiles onto a fine grid around the core, and fitting the reversals with a high order polynomial. The broad distribution encompasses values obtained from modeling efforts \citep{Rubio_da_Costa_2017,Kerr_mgII_1}, and results from the fact that the position of the central reversal evolves in time. We also note that this distribution may be biased to both high and low absolute values. In the former case, since the automatic exposure control of IRIS can only react with a small delay, some of the brightest spectra often saturate the detector, leading to the possibility of the strongest blueshifted reversals being excluded in our statistics. On the other hand, shallow central reversals run the risk of being mislabeled into other groups.

\subsubsection{Information from other spectral windows during the ribbon front phase}
The \ion{Si}{4} line is often assumed to form under optically thin conditions, although this is probably a safe assumption for the quiet Sun, a number of flare observations have hinted at optical depth effects \citep[e.g][]{Tian_2015,Li_2017,Warren_2016}. Furthermore, modeling efforts that include accurate charge exchange processes \citep{Thick_SiIV} show that a substantially larger fraction of \ion{Si}{4} is found at lower temperatures than what is normally predicted by ionization-equilibrium, leading in some cases to optically thick line formation. The complex asymmetric \ion{Si}{4} line profiles associated with the Strong-Ribbon-Front panel of \Cref{probA} contribute to the growing body of evidence for momentary optically thick \ion{Si}{4} line formation in flares. We note that a direct opacity measure can be obtained using the 1394/1403 line ratio, however, the 1394 $\text{\AA}$ line is only available for two of our observations (flare 4 and 14 in \Cref{obs}), and unfortunately is always overexposed for Strong-Ribbon-Front pixels.

We also note that the coronal \ion{Fe}{21} line is not visible during the initial stages of the flare, but can become visible together with some weak ribbon flare spectra. This is in agreement with \cite{LoopFootBlueGraham}, who observed a delay time of more than a minute between the signatures of chromospheric condensation and coronal upflow. 

\ion{Fe}{2} being in emission, especially during the strong ribbon front profiles shows that energy is most likely deposited in the low atmosphere, leading to a different temperature gradient. Although the intensity information has been normalized out of the spectra, the emission in the most probable \ion{Fe}{2} spectra corresponding to the \textit{Strong-Ribbon} case, must be far greater than the surrounding continuum emission, which in contrast appears to be flat. Such strong line emission is commonly seen in IRIS flares \citep[e.g.,][]{kleintetal2016, kleintetal2017} and can exceed the Balmer continuum significantly \citep{heinzelkleint2014}. This leads us to believe that for some high energy cases, where the accelerated electron energies exceed $50~\text{keV}$, the lower stationary layers may be heated at a faster rate than the downward moving condensational layer \citep{Adam2}.

For readers interested in these profile types, we suggest the observation covering Flare 12 of Table \ref{obs}. This flare is confined between two sunspots, and as a result, the energy from consecutive reconnections is deposited within a small region of plasma, providing sustained \textit{Ribbon-Front-Spectra}  within the same pixels. This surplus of energy also results in the emission of other metallic lines which appear as spikes along the \ion{Mg}{2} profiles.

\subsection{Single-Peaked-Spectra}

After the passing of a flare ribbon, the chromosphere is compressed and heated. The enhanced temperatures and densities result in the formation of \textit{Single-Peaked} \ion{Mg}{2} spectral types as seen in both the yellow and orange upper panels of \Cref{probA}. The enhanced densities may lead to thermalization lengths that are comparative or typically smaller than the length scales defining local temperature variations, which in turn leads to a temperature sensitivity in the source function of the cores $S(n_e,T_e)$, allowing them to rise with the electron temperature instead of fall off with radial height. Other explanations include a superposition of plasma moving in opposite directions \citep{Rubio_da_Costa_2017}, or strongly enhanced Stark broadening \citep{Zhu_2019}.

The yellow spectral types often occur after the \textit{Ribbon-Front-Spectra} as seen in \Cref{Temporal_evolution}. Because they typically evolve from ribbon profiles, they are extremely broad due to the enhanced thermal motions. \ion{Mg}{2} has redshifted cores, indicating bulk atmospheric downflows. 
The \ion{C}{2} and \ion{Si}{4} windows are still in downflow and retain many of the same features as before, but lack the reversals associated with the \textit{Ribbon-Front} spectra. Interestingly, \ion{Fe}{21} finally appears with some examples of large upflows and becoming mostly stationary, in agreement with the example shown by \citet{battagliaetal2015} who found blueshifts near footpoints and low velocities in the loops afterwards. The \ion{Fe}{2} line is still mostly in emission, with a few pixels showing absorption.

Due to the similarity between the yellow and orange spectral types discussed above, it is hard to impose a strict time ordering, however, in general, yellow spectral types are followed by orange spectral types that have smaller widths and equally intense triplet emission. Notably, in the middle right panel of \Cref{probA}, \ion{Si}{4} still shows strong downflows but has developed a prominent stationary component, while \ion{Fe}{21} shows no signs of upflow profiles, but remains in emission at line center. Finally, \ion{Fe}{2} appears either in emission and absorption.

\subsection{Downflows}
During the gradual phase we observe a large variety of spectral types associated with downflows, two of which can be seen in the bottom panels of \Cref{probA}, with a slightly less extended red wing in the bottom left panel that tends to appear before spectra with substantially extended red wings as seen in the bottom right panel. The fact that the majority of these spectral types occur during the gradual phase as seen in \Cref{Temporal_evolution} and \Cref{temporalevol}, imply that they are mostly associated with coronal rain. When cool condensing material cascades back down the flare loops, the single peaked spectral types that came before, develop a red wing enhancement, which is mirrored both in \ion{C}{2} and \ion{Si}{4}. The velocities are relatively similar at $\approx 100$ km s$^{-1}$ in all spectral lines and supersonic, similar to previous IRIS coronal rain observations \citep[e.g.,][]{kleintetal2014, lacatusetal2017}. \ion{Fe}{21} still appears in emission and stationary, while \ion{Fe}{2} can be either in emission or absorption. When \ion{Mg}{2} displays extremely broad and intense red wing emission (red group), \ion{Fe}{2} appears almost exclusively in absorption, possibly indicating that the lower chromosphere has finally dissipated its energy and cooled. 

\begin{figure*}[t] 
\centering
\includegraphics[trim={0cm 0cm 0cm 0cm},clip,width=.49\textwidth]{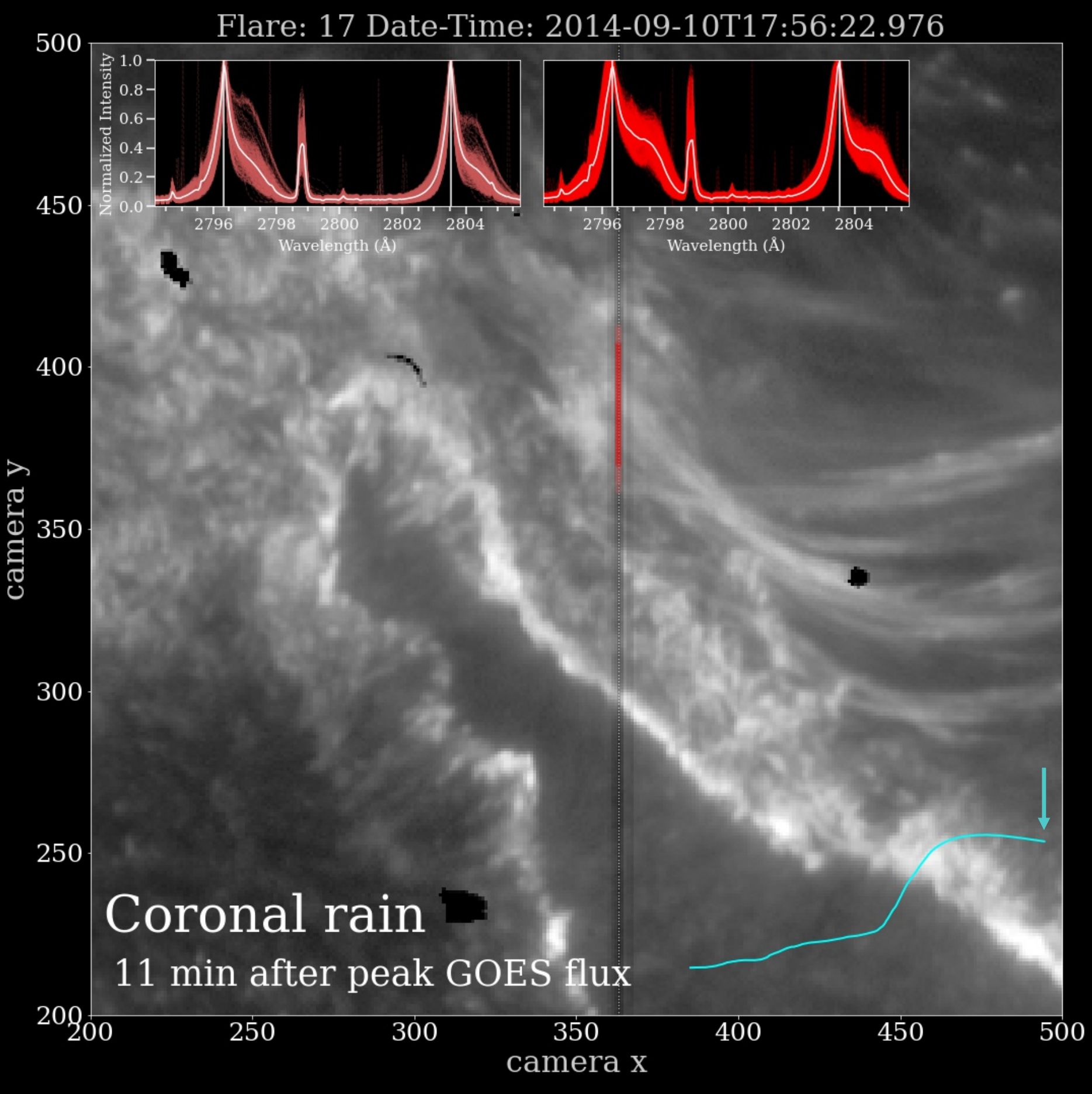} 
\includegraphics[trim={0cm 0cm 0cm 0cm},clip,width=.49\textwidth]{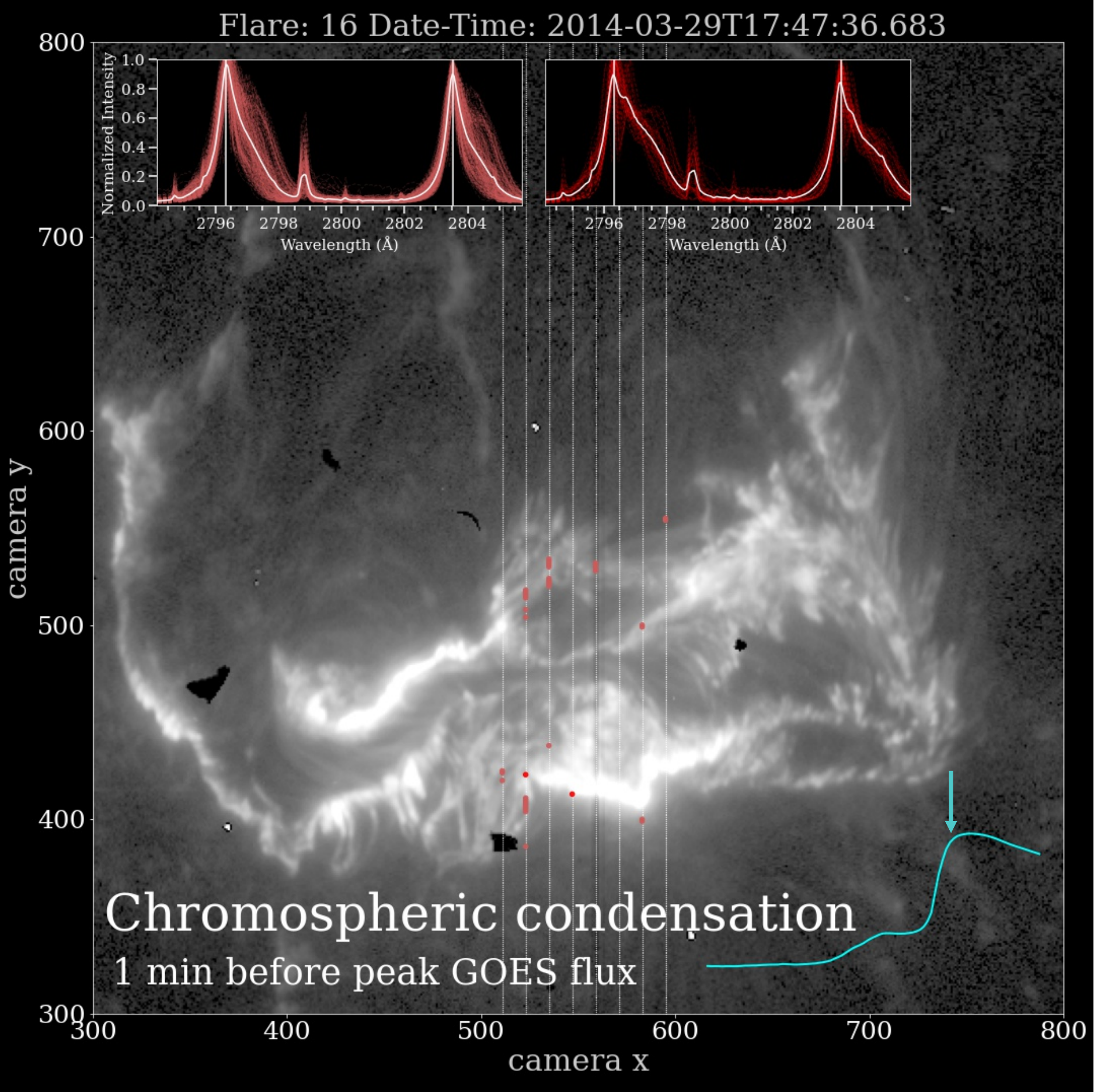}
\caption{Large downflows occur both during the gradual phase as coronal rain and sometimes during the impulsive phase due to chromospheric condensation. The left panel shows a 1400-SJI of flare 17 as the upper portion of IRIS's spectrograph captures down flowing material during the gradual flare phase as indicated by the red highlighted pixels along the slit. The spectra associated with these pixels have been plotted in the upper left two inserts. The first insert shows spectra that have been grouped into the bottom left spectral type of \Cref{probA}, while the second insert is associated with the extended shoulder downflows (bottom right of \Cref{probA}). Similar spectral shapes can be seen in the right panel during the impulsive phase of flare 16. The GOES X-ray curve for each flare is schematically plotted in blue at the bottom of both panels, with an arrow indicating the approximate time each SJI was rendered.}
\label{CR_CC}
\end{figure*}

We analyzed the integrated line ratios for both the core and red wing components of the h\&k-lines by fitting two Gaussian's to each line. We found that on average, the red component is more optically thin than the core, but is still far from the optically thin ratio of 2. We found that the optical depths of these components were independent of raster, meaning opacity within the components do not depend on time. We will investigate the optical thickness of the ratios in more detail in a further study, including the seemingly more pronounced red component of \ion{Si}{4}.\\

Although the large majority of these spectral types can be associated with coronal rain and occur during the gradual flare phase, very similar-looking spectra can also occur during the impulsive phase due to chromospheric condensation \citep[e.g.,][]{Brannon_2016,multi_line_flare}. Figure \ref{CR_CC} shows the location of \ion{Mg}{2} spectra with large red wing enhancements during the gradual phase of flare 17 and the impulsive phase of flare 16. The spectra in the upper left-hand corner of each panel have been automatically found by our algorithm at the red locations indicated along the slit. The two inserts of spectra in both images are associated with the bottom two spectral types of \Cref{probA}. Although the spectra from both flares are similar, there appear to be some subtle differences. The red wing triplet emission is significantly enhanced in relation to the core for the coronal rain, while the peaks of the line cores are broader for the case of the condensation, possibly due to the increased turbulence and energy from the non-thermal electrons. Additionally, the right insert for the coronal rain shows a distinct red wing shoulder, indicating a superposition between a stationary component at line-center and a downflowing component, contrary to the right insert for the condensation, which appears to show spectra with a flatter red wing runoff.

\section{Conclusion and outlook}

A trained Mutual Information Neural Estimator can be used to calculate a tight lower bound on the degree to which any two lines within a single IRIS pixel are correlated. High correlations imply the possibility of correctly guessing the shape of line $\mathcal{L}1$ from the shape of $\mathcal{L}2$. Additionally, the $k$-means clustering algorithm can be used to calculate simple probability distributions by grouping similar but non-identical spectra together, and counting the number of co-occurrences of each spectral type within the same pixel. These simple models can then be used to plot the most and least likely responses of line $\mathcal{L}1$ given a particular spectral type from $\mathcal{L}2$. These models are based on a minimal set of assumptions, namely that the number of spectral types is sufficient to capture the large variety of spectral shapes encountered during a solar flare. The methods developed here are not limited to the windows or observations encountered in this paper, and can be readily applied to any number of spectral lines and instruments. For instance, one could analyze particular responses to \ion{O}{4} instead of \ion{Mg}{2}, provided there are enough IRIS observations available to build reliable probability distributions. We used these methods to analyze a set of typical \ion{Mg}{2} flare spectra and their accompanying responses over 4 different spectral windows, finding the following results:

\begin{itemize}
\item The correlations between line shapes are maximized over the flare ribbons, meaning that in these highly energetic regions of the atmosphere, it is possible to predict the emission of a particular ion based on the emission from a different ion.
\item The blue-shifted reversal in \ion{Mg}{2} associated with \textit{Ribbon-Front-Spectra} also typically co-occur both in \ion{C}{2} and \ion{Si}{4}, although less often. This reversal is likely due to downflows and coherent scattering may not be necessary for its formation, especially in lines with less optical depth than \ion{Mg}{2}.
\item \ion{Si}{4} appears to show optical depth effects during the impulsive phase at the edge of flare ribbons.
\item \ion{C}{2} and \ion{Si}{4} appear to be in downflow for the entirety of each flare observation, with prominent downflows even in \ion{Fe}{2} during the initial stages of the flare.
\item The emission of the \ion{Fe}{2} line indicates deep atmospheric heating which gradually decrease over the duration of the flare.
\item \ion{Fe}{21} only goes into emission once the flare ribbons have passed over a particular region, showing a delay time between downflows in the chromosphere and large coronal upflows, as also found by \citet{LoopFootBlueGraham}. \ion{Fe}{21} is then seen to be in emission and stationary for the remaining flare time.
\item Our analysis shows that in general, \ion{Mg}{2} spectra located over flare ribbons typically develop a blue shifted reversal which initially migrates away from line center before returning to its rest wavelengths. These spectra are accompanied by strong triplet-line emission, and transition into single peaked shapes while still retaining the large widths associated with the \textit{Ribbon-Front-Spectra}. The line widths slowly erode over time, while the triplet emission remains strong , but declines in the gradual flare phase. During the gradual phase, strong red-enhancements indicating coronal rain are prominent.
\item The ratios associated with the downflowing spectra indicate optically thick line formation both for the core and red wing component. The line ratios appear to be invariant in time.
\end{itemize}

It is important to keep in mind that these results are based on statistics derived from hundreds of millions of IRIS spectra, and that the relationships between different spectral lines shown here, represent the aggregate behaviour of 21 large solar flares. Our method is well-suited to analyze large data sets efficiently and could in the future be very useful for the analysis of DKIST observations, which are expected to exceed Petabytes of data per year \citep{dkist2020}. By analyzing overview images, such as Figure \ref{probA}, one can immediately determine the applicability of such multi-line analyses, as each panel represents the only possible set of cotemporal responses, independently of the sample size.

The MINE-network indicates that it should be possible (for instance in the case of (\ion{Mg}{2}|\ion{C}{2})) to accurately reconstruct spectral data from one line based on the information of another. These methods represent a first approximation to achieving this goal, and would be greatly enhanced via the use of a continuous generative model which takes into account the temporal domain. The ability to predict missing data could be particularly beneficial if two instruments ran at different cadences, or in the case of overexposure.

\acknowledgements
For the derivation of all mutual information quantities, we used the deep learning library PyTorch \citep{PyTorch}, as well as the Scikit-Learn module for all implementations of the $k$-means algorithm \citep{SK}. The pre-processing was accomplished with IRISreader, a library specifically developed for handling large volumes of IRIS data \citep{IRISreader}. We would like to thank the Swiss National Science Foundation for funding this research under grant number 407540\_167158 and a PRIMA grant, as well as LMSAL and NASA for allowing us to download all the IRIS data from their servers. IRIS is a NASA small explorer mission developed and operated by LMSAL with mission operations executed at NASA Ames Research center and major contributions to downlink communications funded by ESA and the Norwegian Space Centre. We thank S. Voloshynovskiy, S. Krucker, D. Ulmann, C. Huwyler, and M. Melchior for discussions.

\bibliographystyle{apj}
\bibliography{journals,references}

\end{document}